\documentclass[twocolumn,pra,showpacs,superscriptaddress]{revtex4}
\usepackage{amssymb}
\usepackage{amsmath}
\usepackage{graphicx}
\usepackage{subfigure}
\usepackage{natbib}
\usepackage{epsfig}
\usepackage{amsfonts}
\usepackage{mathrsfs}
\usepackage{CJK}
\usepackage{color}
\usepackage[toc,page,title,titletoc,header]{appendix}

\begin{document}

\title{Interatomic collisions in two-dimensional and quasi-two-dimensional
confinements with spin-orbit coupling}
\author{Peng Zhang}
\affiliation{Department of Physics, Renmin University of China, Beijing, 100190, China}
\author{Long Zhang}
\affiliation{Hefei National Laboratory for Physical sciences at Microscale, University of
Science and Technology of China, Hefei, Anhui 230027, China}
\author{Wei Zhang*}
\affiliation{Department of Physics, Renmin University of China, Beijing, 100190, China}

\begin{abstract}
We investigate the low-energy scattering and bound states of two
two-component fermionic atoms in pure two-dimensional (2D) and quasi-2D
confinements with Rashba spin-orbit coupling (SOC). We find that the SOC
qualitatively changes the behavior of the 2D scattering amplitude in the
low-energy limit. For quasi-2D systems we obtain the analytic expression for
the effective-2D scattering amplitude and the algebraic equations for the
two-atom bound state energy. We further derive the effective 2D contact
potential for the interaction between ultracold atoms in the quasi-2D
confinement. This effective potential can be used in the research on
many-body physics of quasi-2D ultracold fermi gases with Rashba SOC.
\end{abstract}

\maketitle

\section{Introduction}

The discussion on synthetic gauge field~\cite{lin-09a} and spin-orbit
coupling (SOC)~\cite{SOC} in bosonic and fermionic systems has recently
drawn great attention~\cite{zhai-11} since its experimental realization in
cold atomic gases. By applying counter propagating Raman pulses with tunable
properties, the effective gauge field and SOC have been accomplished in
ultra-cold gases of both bosons~\cite%
{NIST,NIST_elec,SOC,collective_SOC,NIST_partial,ourdecay,JingPRA} and
fermions~\cite{SOC_Fermi,SOC_MIT}. These experimental achievements add an
additional piece to the already versatile toolbox of manipulation of cold
atoms, and may help paving a new route toward the realization and
investigation on novel quantum states. A considerable amount of theoretical
interest has been stimulated in the understanding of the SOC effect in both
bosonic~\cite{wu-11, wang-10, ho-11, xu-11, xu-11b, zhou-11, sinha-11,
hu-12,cui-12} and fermionic~\cite{cui-12,vyasanakere-11, gong-11, yu-11,
hu-11, iskin-11, yi-11, han-12, dellanna-11, zhou-11b, chen-12, seo-11,
huang-11, he-11,Takei-11} systems.

Among these works, one important direction is the study of SOC effect in low
dimensionality where many interesting novel quantum states may be present.
In particular, a BEC with a half-quantum-angular-momentum vortex may exist
in two-dimensional (2D) bosonic system with Rashba SOC~\cite{zhou-11,
sinha-11, hu-12}. In two-component Fermi gases, a topological superfluid is
proposed in 2D configuration, and can supports zero-energy Majorana modes
which are related to fault tolerant quantum computation~\cite{zhou-11b,
chen-12}.

In realistic experiments of ultra-cold atoms, the low-dimensional physics
are usually studied in a quasi-low-dimensional configuration, where atoms
are strongly confined in one or two spatial dimensions such that the
trapping frequencies along these directions dominate all other relevant
energy scales in an effective low-dimensional Hamiltonian. In the energy
scale of interest, this effective Hamiltonian should catch the same physics
as the original Hamiltonian, which usually corresponds to two-body processes
in the context of cold atoms as the particle separation is much larger than
the range of interatomic interaction. Thus, to write down the correct
effective low-dimensional Hamiltonian, one needs to investigate the two-body
physics in the quasi-low-dimensional confinement, and express the effective
low-dimensional scattering amplitude or the two-body binding energy in terms
of the \textquotedblleft control parameters", including the
three-dimensional (3D) scattering length and the intensity of the
confinement. In the absence of SOC, this analysis has been performed for
quasi-one-dimensional (1D)~\cite{olshanii-98,Duan-07} and quasi-2D~\cite%
{petrov01, Duan-07} configurations, and leads to the unique feature of
confinement-induced resonance~\cite{olshanii-98} and a different
renormalization scheme around Feshbach resonances~\cite{Duan-07,zhang-08}.
In the presence of SOC, a discussion of two-body physics within
quasi-low-dimensional confinement and a derivation of the effective
low-dimensional Hamiltonian is still lack.

In this paper, we investigate two-body scattering process and bound state
energy of two-component fermionic atoms in 2D and quasi-2D geometries with a
Rashba type SOC. For 2D systems, we find that when the total momentum of two
atoms is zero, the 2D inter-atomic scattering amplitude in the low-energy
limit will be qualitatively altered by the presence of SOC, rendering a
\textit{polynomial} rather than logarithmical decay to zero. For quasi-2D
systems with a strong harmonic confinement along the axial $z$-direction, we
obtain analytic expression for the effective 2D scattering amplitude. We
also derive the equation for the binding energy of two-atom bound states
(dimer), as well as the effective mass of the dimer. We find that the
presence of Rashba SOC tends to \textit{enhance} the two-body binding
energy. This observation can be qualitatively understood by noticing that
the density of states in the low energy limit is increased as the ground
state becomes degenerate with SOC. Based on these results, we further derive
an effective 2D interaction between atoms in the quasi-2D gas with Rashba
SOC, and map out two effective 2D Hamiltonians which are responsive for
energy regimes around the two-body binding energy and close to the
single-particle threshold. These effective models can be used to study
many-body properties of quasi-2D gas with attractive or repulsive
interactions, respectively. Our results provide the possibility to control
the effective 2D physics via 3D parameters including the scattering length
and the intensity of $z$-confinement. The method developed in this
manuscript can be directly generalized to other systems with all kinds of
atoms and arbitrary types of SOC.

The remainder of this manuscript is organized as follows. In Sec. II, we
discuss the two-body physics in 2D. In Sec. III, we calculate the low-energy
scattering amplitude and the binding energy of dimers in a quasi-2D geometry
with a strong harmonic trapping potential along the axial $z$-direction. By
matching the two-body physics, we construct in Sec. IV the effective 2D
Hamiltonian which can describe the low-energy behavior of the quasi-2D gas.
The main results are discussed and summarized in Sec. V, while some details
of our calculations are explained in the appendixes.


\section{2D scattering with Rashba SOC}

In this section we discuss the scattering and bound states of two spin-$1/2$
fermionic atoms in a pure-2D geometry ($x$-$y$ plane) with Rashba SOC. In
such a system, the total momentum $\boldsymbol{q}$ of the two atoms is
conserved. Then the spatial motion of the mass center of the two atoms is
separated from the relative motion and the spin of the two atoms. Thus, the
quantum state of the two-atom relative motion can be described by a spinor
wave function
\begin{eqnarray}
|\psi \left( \boldsymbol{\rho }\right) \rangle &=&\psi _{\uparrow \uparrow
}\left( \boldsymbol{\rho }\right) |\uparrow \rangle _{1}|\uparrow \rangle
_{2}+\psi _{\uparrow \downarrow }\left( \boldsymbol{\rho }\right) |\uparrow
\rangle _{1}|\downarrow \rangle _{2}  \notag \\
&&+\psi _{\downarrow \uparrow }\left( \boldsymbol{\rho }\right) |\downarrow
\rangle _{1}|\uparrow \rangle _{2}+\psi _{\downarrow \downarrow }\left(
\boldsymbol{\rho }\right) |\downarrow \rangle _{1}|\downarrow \rangle _{2}
\end{eqnarray}%
with $\boldsymbol{\rho =}\left( x,y\right) $ the 2D relative coordinate of
the two atoms and $|\uparrow \rangle _{1\left( 2\right) }$ and $|\downarrow
\rangle _{1\left( 2\right) }$ the spin eigen-states of the 1st (2nd) atom.
It is apparently that the wave function $|\psi \left( \boldsymbol{\rho }%
\right) \rangle $ can also be considered as a $\boldsymbol{\rho }$-dependent
spin stae of the two atoms. In this paper, we use the nature unit $\hslash
=m=1$ with $m$ the single-atom mass.

Due to the presence of SOC, the relative motion of the two atoms depends on
their center-of-mass momentum $\boldsymbol{q}$. For a given value of $%
\boldsymbol{q=}\left( q_{x},q_{y}\right) $, the Hamiltonian for atomic
relative motion and spin degrees of freedom is given by
\begin{equation}
H^{\left( \mathrm{2D}\right) }=H_{0}^{\left( \mathrm{2D}\right) }+V_{\mathrm{%
2D}}\left( \rho \right) ,  \label{h2d}
\end{equation}%
where the 2D free Hamiltonian $H_{0}^{\left( \mathrm{2D}\right) }$ is given
by%
\begin{eqnarray}
H_{0}^{\left( \mathrm{2D}\right) } &=&-\sum_{\beta =x,y}\frac{\partial ^{2}}{%
\partial \beta ^{2}}+\frac{\xi }{2}\sum_{j=1,2}\left[ \frac{q_{x}}{2}+\left(
-1\right) ^{j}p_{x}\right] \hat{\sigma}_{x}^{\left( j\right) }  \notag \\
&&+\frac{\xi }{2}\sum_{j=1,2}\left[ \frac{q_{y}}{2}+\left( -1\right)
^{j}p_{y}\right] \hat{\sigma}_{y}^{\left( j\right) }  \label{h0soc}
\end{eqnarray}%
with $\boldsymbol{p}=\left( p_{x},p_{y}\right) $ the relative momentum of
the two atoms. The Rashba SOC is described by the second term of Eq.~(\ref%
{h0soc}), where $\boldsymbol{q}/2+\left( -1\right) ^{j}\boldsymbol{p}$ is
the momentum of the $j$-th atom. The spin operator $\hat{\sigma}%
_{x,y}^{\left( j\right) }$ for the $j$-th atom is defined as $\hat{\sigma}%
_{x}^{\left( j\right) }=|\uparrow \rangle _{j}\langle \downarrow |+|\uparrow
\rangle _{j}\langle \downarrow |$ and $\hat{\sigma}_{y}^{\left( j\right)
}=-i|\uparrow \rangle _{j}\langle \downarrow |+i|\downarrow \rangle
_{j}\langle \uparrow |$. Without loss of generality, we assume the SOC
intensity $\xi $ in Eq.~(\ref{h0soc}) is a positive real number.

In Eq.~(\ref{h2d}) $V_{\mathrm{2D}}\left( \rho \right) $ is the atom-atom
interaction potential with $\rho =\left\vert \boldsymbol{\rho }\right\vert $%
. We assume the effective range of the potential $V_{\mathrm{2D}}\left( \rho
\right) $ is $\rho _{\ast }$, that is the potential becomes negligible in
the region $\rho \gtrsim \rho _{\ast }$. We further assume the SOC is weak
enough so that $\xi \ll 4/\rho _{\ast }.$

To investigate the scattering process in our systems, we first define
single-atom spin state $|\alpha _{j},\boldsymbol{t}\rangle _{j}$ for the $j$%
-th atom as%
\begin{equation}
\frac{\xi }{2}\left( t_{x}\hat{\sigma}_{x}^{\left( j\right) }+t_{y}\hat{%
\sigma}_{y}^{\left( j\right) }\right) |\alpha _{j},\boldsymbol{t}\rangle
_{j}=\alpha _{j}\frac{\xi |\boldsymbol{t}|}{2}|\alpha _{j},\boldsymbol{t}%
\rangle _{j}
\end{equation}%
with $\boldsymbol{t}=(t_{x},t_{y})$ any 2D vector and $\alpha _{j}=\pm 1$.
We further define the two-atom spin state $|\boldsymbol{\alpha }\left(
\boldsymbol{q},\boldsymbol{k}\right) \rangle $ as
\begin{equation}
|\boldsymbol{\alpha }\left( \boldsymbol{q},\boldsymbol{k}\right) \rangle
=|\alpha _{1},\frac{\boldsymbol{q}}{2}+\boldsymbol{k}\rangle _{1}|\alpha
_{2},\frac{\boldsymbol{q}}{2}-\boldsymbol{k}\rangle _{2},
\end{equation}%
with $\boldsymbol{\alpha }=\left( \alpha _{1},\alpha _{2}\right) $ and $%
\boldsymbol{\bar{\alpha}}=\left( \alpha _{2},\alpha _{1}\right) $.

In the scattering process, the incident wave function shoud be the
eigen-state of the Hamiltonian $H_{0}^{\left( \mathrm{2D}\right) }$ for the
free motion of the two fermionic atoms. The straightforward calculation
shows that the such a incident wave function takes the form%
\begin{equation}
|\psi _{c}^{\left( 0\right) }\left( \boldsymbol{\rho }\right) \rangle \!=\!%
\frac{e^{i\boldsymbol{k}\cdot \boldsymbol{\rho }}}{2^{3/2}\pi }|\boldsymbol{%
\alpha }\left( \boldsymbol{q},\!\boldsymbol{k}\right) \rangle \!-\!\frac{%
e^{-i\boldsymbol{k}\cdot \boldsymbol{\rho }}}{2^{3/2}\pi }|\boldsymbol{\bar{%
\alpha}}\left( \boldsymbol{q},\!-\boldsymbol{k}\right) \rangle \!.
\label{psi2db}
\end{equation}%
with incident momentum $\boldsymbol{k}=\left( k_{x},k_{y}\right) $. In this
paper we denote
\begin{equation}
c=\left( \boldsymbol{\alpha },\boldsymbol{q},\boldsymbol{k}\right)
\end{equation}%
as the set of all the three quantum numbers. Here we have considered the
Pauli's principle for the fermonic atoms. The eigen-energy of $H_{0}^{\left(
\mathrm{2D}\right) }$ with respect to the eigen-state $|\psi _{c}^{\left(
0\right) }\left( \boldsymbol{\rho }\right) \rangle $ is
\begin{equation}
\varepsilon _{c}=k^{2}+\frac{\xi }{2}\left( \alpha _{1}\left\vert \frac{%
\boldsymbol{q}}{2}+\boldsymbol{k}\right\vert +\alpha _{2}\left\vert \frac{%
\boldsymbol{q}}{2}-\boldsymbol{k}\right\vert \right)  \label{div}
\end{equation}%
with $k=\left\vert \boldsymbol{k}\right\vert $. Notice that in the presence
of SOC, the scattering threshold, or the minimum value of $\varepsilon _{c}$
with respect to a fixed $\boldsymbol{q}$, is shifted from $0$ to $%
\varepsilon _{\mathrm{thre}}\left( q\right) $ which is given by%
\begin{equation}
\varepsilon _{\mathrm{thre}}\left( q\right) =\left\{
\begin{array}{ll}
-q^{2}/4-\xi ^{2}/4 & (q<\xi ) \\
&  \\
-q\xi /2 & (q>\xi )%
\end{array}%
\right.
\end{equation}%
with $q=|\boldsymbol{q}|$.

Now we consider the scattering state $|\psi _{c}^{\left( +\right) }\left(
\boldsymbol{\rho }\right) \rangle $ with respect to the incident state $%
|\psi _{c}^{\left( 0\right) }\left( \boldsymbol{\rho }\right) \rangle $. We
assume the scattering energy $\varepsilon _{c}$ is low enough with $k\ll
1/\rho _{\ast }$. In such a low-energy case and within the region of $\rho
\gtrsim \rho _{\ast }$, the wave function of the scattering state $|\psi
_{c}^{\left( +\right) }\left( \boldsymbol{\rho }\right) \rangle $ can be
expressed as (see discussion in Appendix A and in Ref.~\cite{petrov01})
\begin{equation}
|\psi _{c}^{\left( +\right) }\left( \boldsymbol{\rho }\right) \rangle
\approx |\psi _{c}^{\left( 0\right) }\left( \boldsymbol{\rho }\right)
\rangle \!+A\left( c\right) g\left( \varepsilon _{c};\boldsymbol{\rho },%
\boldsymbol{0}\right) |0,0\rangle ,  \label{sw2}
\end{equation}%
where
\begin{equation}
|0,0\rangle =1/\sqrt{2}\left( |\uparrow \rangle _{1}|\downarrow \rangle
_{2}-|\downarrow \rangle _{1}|\uparrow \rangle _{2}\right)
\end{equation}
is the spin singlet state, and the 2D free Green's function $g\left( \eta ;%
\boldsymbol{\rho },\boldsymbol{\rho }^{\prime }\right) $ is given by
\begin{equation}
g\left( \eta ;\boldsymbol{\rho },\boldsymbol{\rho }^{\prime }\right) =\frac{1%
}{\eta +i0^{+}\!-\!H_{0}^{\left( \mathrm{2D}\right) }\!}\delta \left(
\boldsymbol{\rho }-\boldsymbol{\rho }^{\prime }\right) ,  \label{g2d}
\end{equation}%
and can be considered as a $\left( \boldsymbol{\rho },\boldsymbol{\rho }%
^{\prime }\right) $-dependent operator for the two-atom spin.

The coefficient $A\left( c\right) $ in Eq.~(\ref{sw2}) can be derived with
the following analysis. First, it can be proved (see Appendix B) that in the
\textit{small-distance region} $\rho _{\ast }\lesssim \rho \ll 1/k$ the
function $|\psi ^{\left( c,+\right) }\left( \boldsymbol{\rho }\right)
\rangle $ behaves as
\begin{equation}
|\psi ^{\left( c,+\right) }\left( \boldsymbol{\rho }\right) \rangle \propto
\left( \ln \rho -\ln d\right) |0,0\rangle  \label{src2d}
\end{equation}%
Here, the $\boldsymbol{\rho }$-independent factor $\ln d$ is determined by $%
\xi $ and the detail of the potential $V_{\mathrm{2D}}\left( \rho \right) $
(Appendix B). and almost independent on the scattering energy $\varepsilon
_{c}$ in the low-energy case~\cite{petrov01}. Second, the calculations in
Appendix C shows that, in the small-distance region we have%
\begin{eqnarray}
\langle 0,0|g\left( \eta ;\boldsymbol{\rho },\boldsymbol{0}\right)
|0,0\rangle &\approx &\frac{1}{2\pi }\left[ \ln \rho \!+\!C+\!\ln \left( -i%
\frac{\sqrt{\varepsilon _{c}}}{2}\right) \right]  \notag \\
&&+\lambda \left( \varepsilon _{c},\boldsymbol{q}\right)  \label{gg}
\end{eqnarray}%
where $C=0.5772...$ is the Euler gamma number and the function $\lambda
(\eta ,\boldsymbol{q})$ is defined as%
\begin{eqnarray}
&&\lambda (\eta ,\boldsymbol{q})=\frac{1}{\left( 2\pi \right) ^{2}}\sum_{%
\boldsymbol{\alpha }^{\prime \prime }}\int d\boldsymbol{k}^{\prime \prime
}\left\vert \langle 00|\boldsymbol{\alpha }^{\prime \prime }(\boldsymbol{q},%
\boldsymbol{k}^{\prime \prime })\rangle \right\vert ^{2}\times  \notag \\
&&\left( \frac{1}{\eta +i0^{+}-\varepsilon _{c^{\prime \prime }}}-\frac{1}{%
\eta +i0^{+}-|\boldsymbol{k}^{\prime \prime }|^{2}}\right) ,  \label{lam2d}
\end{eqnarray}%
with $c^{\prime \prime }=(\boldsymbol{\alpha }^{\prime \prime };\boldsymbol{q%
};\boldsymbol{k}^{\prime \prime })$. For $\boldsymbol{q}=0$, a
straightforward calculation shows that the function $\lambda (\eta ,%
\boldsymbol{0})$ can be further simplified as
\begin{eqnarray}
&&\lambda (\eta ,\boldsymbol{0})=\frac{1}{2\left( 2\pi \right) }\times
\notag \\
&&\left\{
\begin{array}{l}
\frac{-\xi }{2\sqrt{-4\eta -\xi ^{2}}}\left( \pi +2\arctan \left[ \frac{\xi
^{2}+2\eta }{\xi \sqrt{-4\eta -\xi ^{2}}}\right] \right) ;(\eta <-\frac{\xi
^{2}}{4}) \\
\\
\frac{-i\pi \xi }{2\sqrt{\eta +\xi ^{2}/4}}+\xi \frac{\mathrm{arctanh}\left(
2\sqrt{\eta +\xi ^{2}/4}/\xi \right) }{\sqrt{\eta +\xi ^{2}/4}};(-\frac{\xi
^{2}}{4}<\eta <0) \\
\\
\frac{\xi }{\sqrt{\eta +\xi ^{2}/4}}\mathrm{arctanh}\frac{\xi }{2\sqrt{\eta
+\xi ^{2}/4}};(0<\eta <1-\frac{\xi ^{2}}{4})%
\end{array}%
\right. .  \notag \\
&&  \label{g00}
\end{eqnarray}

Substituting Eq.~(\ref{gg}) into (\ref{sw2}), we get the expression for $%
|\psi _{c}^{\left( +\right) }\left( \boldsymbol{\rho }\right) \rangle $ in
the small-distance region
\begin{eqnarray}
&&|\psi _{c}^{\left( +\right) }\left( \boldsymbol{\rho }\right) \rangle
\approx |\psi _{c}^{\left( 0\right) }\left( \boldsymbol{\rho }\right)
\rangle +  \notag  \label{sea} \\
&&\!\frac{A\left( c\right) }{2\pi }\left\{ \ln \rho \!+\!C+\!\ln \left( -%
\frac{i\sqrt{\varepsilon _{c}}}{2}\right) \!+\!\lambda \left[ \varepsilon
_{c},\boldsymbol{q}\right] \right\} |0,0\rangle .
\end{eqnarray}%
Comparing Eqs.~(\ref{sea}) and (\ref{src2d}), we obtain the expression for
the parameter $A\left( c\right) $:%
\begin{equation}
A\left( c\right) =\frac{(2\pi )\!\langle 0,0|\psi _{c}^{\left( 0\right)
}\left( \boldsymbol{0}\right) \rangle }{i\pi /2-C-\ln \left( d\sqrt{%
\!\varepsilon _{c}}/2\right) -\left( 2\pi \right) \!\lambda \left(
\varepsilon _{c},\boldsymbol{q}\right) }.  \label{a2d}
\end{equation}%
According to Eq.~(\ref{sw2}), $A\left( c\right) $ completely determines the
behavior of the scattering-state wave function $|\psi ^{\left( c,+\right)
}\left( \boldsymbol{\rho }\right) \rangle $ in the region of $\rho \gtrsim
\rho _{\ast }$.

According to scattering theory~\cite{taylor}, we can define the 2D
scattering amplitude $f^{\left( \mathrm{2D}\right) }$ between the incident
state $|\psi _{c}^{\left( 0\right) }\left( \boldsymbol{\rho }\right) \rangle
$ and an energy-conserved output state $|\psi _{c^{\prime }}^{\left(
0\right) }\left( \boldsymbol{\rho }\right) \rangle $ with $\boldsymbol{q}%
^{\prime }=\boldsymbol{q}$ and $\varepsilon _{c}=\varepsilon _{c^{\prime }}$
as
\begin{equation}
f^{\left( \mathrm{2D}\right) }\left( c^{\prime }\leftarrow c\right) =-2\pi
^{2}\int d\boldsymbol{\rho }\langle \psi _{c^{\prime }}^{\left( 0\right)
}\left( \boldsymbol{\rho }\right) |V_{\mathrm{2D}}|\psi _{c}^{\left(
+\right) }\left( \boldsymbol{\rho }\right) \rangle .
\end{equation}%
A straightforward calculation (Appendix D) shows that we can express $%
f^{\left( \mathrm{2D}\right) }$ in terms of the coefficient $A\left(
c\right) $
\begin{equation}
f^{\left( \mathrm{2D}\right) }\left( c^{\prime }\leftarrow c\right) =-2\pi
^{2}\!\langle \psi _{c^{\prime }}^{\left( 0\right) }\left( \boldsymbol{0}%
\right) |0,0\rangle A\left( c\right) .  \label{t2}
\end{equation}

Now we discuss the two-body bound states with Rashba SOC in two-dimensions,
as is also investigated in Ref. \cite{Takei-11}. As shown in Appendix A,
when the energy $\varepsilon _{b}$ of the bound state is close enough to the
scattering threshold $\varepsilon _{\mathrm{thre}}(q)$, or the condition $%
\varepsilon _{\mathrm{thre}}(q)-\varepsilon _{b}\ll 1/\rho _{\ast }^{2}$ is
satisfied, the wave function $|\psi _{b}\left( \boldsymbol{\rho }\right)
\rangle $ of the two-atom bound state can be approximated as%
\begin{equation}
|\psi _{b}\left( \boldsymbol{\rho }\right) \rangle \approx Bg\left(
\varepsilon _{b};\boldsymbol{\rho },\boldsymbol{0}\right) |0,0\rangle
\label{psib2d}
\end{equation}%
in the region $\rho \gtrsim \rho _{\ast }$. Here, $B$ is the normalization
coefficient and $g$ is defined in Eq.~(\ref{g2d}). The energy $\varepsilon
_{b}$ of the bound state is determined by the condition (appendix B)
\begin{eqnarray}
\langle 0,0|\psi _{b}\left( \boldsymbol{\rho }\right) \rangle \propto \ln
\rho -\ln d  \label{bb1}
\end{eqnarray}
in the region $\rho _{\ast }\lesssim \rho \ll 1/\sqrt{\left\vert \varepsilon
_{b}-\varepsilon _{\mathrm{thre}}\left( q\right) \right\vert }$, or by the
equation
\begin{equation}
-\ln d=C+\ln \left( -\frac{i\sqrt{\varepsilon _{b}}}{2}\right) +\left( 2\pi
\right) \lambda (\varepsilon _{b},\boldsymbol{q}).  \label{eb2d}
\end{equation}%
Here, we use the fact that in the small-distance region the function $%
\langle 0,0|g\left( \varepsilon _{b};\boldsymbol{\rho },\boldsymbol{0}%
\right) |0,0\rangle $ also takes the form as in Eq.~(\ref{gg}), with $%
\varepsilon _{c}$ replaced by the new variable $\varepsilon _{b}$.
Therefore, the bound-state energy $\varepsilon _{b}$ is a function of both
the characteristic length $d$ and the center-of-mass momentum $\boldsymbol{q}
$.

In the discussion above, we obtain the analytical expressions for the
scattering amplitude $f^{\left( \mathrm{2D}\right) }$ and the equation for
the bound-state energy in a pure 2D geometry with Rashba SOC. Comparing our
results with the 2D scattering theory without SOC (see Appendix E and Ref.
\cite{petrov01}), we observe the following two qualitative differences:

First, when the total momentum $\boldsymbol{q}$ of the two atoms is zero,
the SOC changes the dependence of the scattering amplitudes on the
scattering energy $\varepsilon _{c}$. As shown in Appendix E, when there is
no SOC, the scattering amplitude $f_{0}^{\left( \mathrm{2D}\right) }$ decays
to zero logarithmically in the limit $\varepsilon _{c}\rightarrow 0$,
\begin{equation}
\lim_{\varepsilon _{c}\rightarrow 0}f_{0}^{\left( \mathrm{2D}\right)
}\propto \frac{1}{\ln \varepsilon _{c}}.
\end{equation}%
When $\boldsymbol{q}=0$, a Rashba SOC will change the scattering amplitude
through the function $\lambda \left( \varepsilon _{c},\boldsymbol{0}\right) $
in $A\left( c\right) $ and the factor $\langle \psi _{c^{\prime }}^{\left(
0\right) }\left( \boldsymbol{0}\right) |0,0\rangle \!\langle 0,0|\psi
_{c}^{\left( 0\right) }\left( \boldsymbol{0}\right) \rangle $. In
particular, the $\lambda $-function removes the logarithmic behavior of the
scattering amplitude in the region around $\varepsilon _{c}=0$, leading to
\begin{equation}
\lim_{\varepsilon _{c}\rightarrow -\xi ^{2}/4}f^{(\mathrm{2D})}\propto \sqrt{%
\varepsilon _{c}+\frac{\xi ^{2}}{4}}.
\end{equation}%
Namely, in the presence of Rashba SOC, the scattering amplitude $f^{(\mathrm{%
2D})}$ polynomially decays to zero in the limit of $\varepsilon
_{c}\rightarrow -\xi ^{2}/4$.

\begin{figure}[tbp]
\includegraphics[bb=7bp 412bp 314bp 665bp,clip,width=6.5cm]{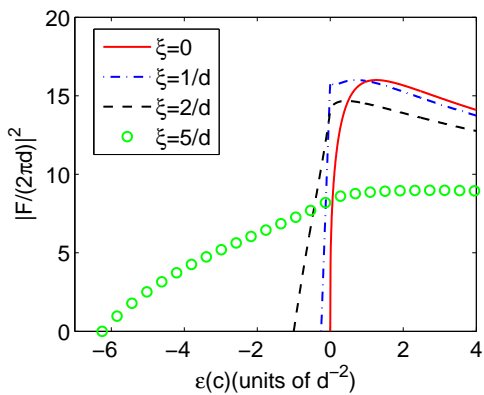}
\caption{(Color online) The variation of function $|F|^{2}$ with 2D
scattering energy $\protect\varepsilon (c)$. Here, $F$ is defined in Eq.~(%
\protect\ref{bigf}) and $d$ is the parameter in Eq.~(\protect\ref{src2d}).
We show results in cases with zero center-of-mass momentum $q=0$ and SOC
intensity $\protect\xi =0$ (red solid line), $1/d$ (blue dashed-dotted
line), $2/d$ (black dashed line) and $5/d$ (green open circle).}
\end{figure}

To illustrate the SOC effect to the 2D scattering amplitude, we plot in
Fig.~1 the mode square of the quantity
\begin{equation}
F\equiv \frac{f^{\left( \mathrm{2D}\right) }\left( c^{\prime }\leftarrow
c\right) }{\langle \psi _{c^{\prime }}^{\left( 0\right) }\left( \boldsymbol{0%
}\right) |0,0\rangle \!\langle 0,0|\psi _{c}^{\left( 0\right) }\left(
\boldsymbol{0}\right) \rangle }  \label{bigf}
\end{equation}%
as a function of the scattering energy $\varepsilon _{c}$ with respect to
different SOC intensities. It can be seen clearly that the function $%
\left\vert F\right\vert ^{2}$ linearly decays to zero in the limit of $%
\varepsilon _{c}\rightarrow -\xi ^{2}/4$ in the presence of SOC, and
logarithmically decays to zero when there is no SOC.

We emphasize that, this change is due to the dispersive relation (\ref%
{div}) of $\varepsilon _{c}$. When $\boldsymbol{q}=0$, $\varepsilon _{c}$ is
independent of the direction of $\boldsymbol{k}$, and takes minimum value $%
-\xi ^{2}/4$ when $k=\xi /2$. Namely, we have $\varepsilon _{c}=-\xi ^{2}/4$
for all momenta $\boldsymbol{k}$ in the cirecle of $(k_{x},k_{y})$
plane. Nevertheless, when $\boldsymbol{q}\neq 0$,
or when SOC is not of Rashba type, this character disappears. In these cases $\varepsilon _{c}$
takes minimum value only when $\boldsymbol{k}$ takes one or two certain
values, as in the systems without SOC. Thus, the 2D scattering
amplitude also logarithmically decays to zero when the scattering energy
approaches to the threshold.

Second, when there is no SOC, the scattering amplitude is independent on the
center-of-mass momentum $\boldsymbol{q}$. This feature is qualitatively
altered by the presence of SOC, as can be clearly seen from Eq.~(\ref{h0soc}%
) where the SOC terms couple the relative motion of the two atoms with the
total momentum $\boldsymbol{q}$. As a consequence, the scattering amplitude $%
f^{\left( \mathrm{2D}\right) }$ becomes a function of $\boldsymbol{q}$. The
similar $\boldsymbol{q}$-dependence can also be observed in the bound-state
energy $\varepsilon _{b}$ of the low-energy bound states \cite{Takei-11}.


\section{Scattering in a quasi-2D confinement with Rashba SOC}

In the previous section, we obtain the two-atom scattering amplitude and
bound-state energy in a pure 2D system with Rashba SOC. Our results show
that the SOC qualitatively changes the 2D scattering amplitude. In a
realistic experiment of cold atoms, the 2D condition is usually realized by
applying a tight confinement along one (say $z$) direction, such that the
degrees of freedom of the single-body Hamiltonian along that specific
direction can be approximately integrated out. In this section, we study the
scattering and bound states of two spin-$1/2$ fermionic atoms in a quasi-2D
configuration. We conclude that the qualitative effects introduced by Rashba
SOC to the two-body physics in pure 2D geometry also exist in the quasi-2D
cases. Besides, we show how the factor $\ln d$ in the expressions of 2D
scattering amplitude and bound state energy can be effectively tuned by the
atomic 3D scattering length and the intensity of $z$-confinement.


\subsection{System and Hamiltonian}

We consider a quasi-2D configuration with a harmonic trap with frequency $%
\omega $ applied along the $z$-direction, while the atomic motion in the $x$-%
$y$ plane is free. In the quasi-2D case, the total momentum $\boldsymbol{q}$
of the two atoms in the $x$-$y$ plane is also conserved and serves as a
parameter for the atomic relative motion, as in the pure 2D case. For a
given value of $\boldsymbol{q}$, the Hamiltonian of the atomic relative
motion and spin states takes the form
\begin{equation}
H=H_{0}^{\left( \mathrm{2D}\right) }+H_{z}+V_{\mathrm{3D}}\left( r\right) .
\label{hq2d}
\end{equation}%
Here, $H_{0}^{\left( \mathrm{2D}\right) }$ is defined in Eq.~(\ref{h0soc}),
\begin{equation}
H_{z}=-\frac{\partial ^{2}}{\partial z^{2}}+\frac{\omega ^{2}z^{2}}{4}-\frac{%
\omega }{2}
\end{equation}%
is the Hamiltonian for the two-atom relative motion in the $z$-direction,
with $z$ the relative coordinate of the two atoms in this direction. Note
that we have shifted the zero-energy point for our convenience, since the
harmonic confinement allows us to separate the relative motion in the $z$%
-direction from the center-of-mass degree of freedom.

In Eq.~(\ref{hq2d}), the atom-atom interaction potential $V_{\mathrm{3D}%
}\left( r\right) $ is a function of the relative position $\boldsymbol{r=}%
\left( \boldsymbol{\rho },z\right) =(x,y,z)$ of the two atoms in three
dimensions. Here, we have $r=|\boldsymbol{r}|$ and denote the effective
range of $V_{\mathrm{3D}}\left( r\right) $ as $r_{\ast }$. In the region $%
r\gtrsim r_{\ast }$, the potential $V_{\mathrm{3D}}\left( r\right) $ becomes
negligible. For simplicity, we further assume $V_{\mathrm{3D}}\left(
r\right) $ is independent on the atomic spin, and consider only the weak SOC
case with $4/\xi \gg r_{\ast }$. 

\subsection{Two-atom scattering state}

In the quasi-2D configuration, we consider only the scattering between two
fermionic atoms in the ground state of $H_{z}$. Then the incident wave
function is given by
\begin{equation}
|\Psi _{c}^{\left( 0\right) }\left( \boldsymbol{r}\right) \rangle =\varphi
_{0}\left( z\right) |\psi _{c}^{\left( 0\right) }\left( \boldsymbol{\rho }%
\right) \rangle  \label{isq2d}
\end{equation}%
with $|\psi _{c}^{\left( 0\right) }\left( \boldsymbol{\rho }\right) \rangle $
defined in Eq.~(\ref{psi2db}). Here, $\varphi _{n_{z}}\left( z\right) $ is
the $n_{z}$-th eigen-wave function of $H_{z}$. Without loss of generality,
we set the phase of $\varphi _{n_{z}}\left( z\right) $ so that $\varphi
_{n_{z}}\left( 0\right) $ is real. We further assume the energy gap between
the incident state and the threshold $\varepsilon _{\mathrm{thre}}$ is
smaller than the trapping frequency along the $z$-direction, i.e. $%
|\varepsilon _{c}-\varepsilon _{\mathrm{thre}}(q)|<\omega $. In this case,
the energy-conserved output states of the scattering are also in the ground
transverse channel with $\varphi _{0}\left( z\right) $.

Next, we calculate the scattering state $|\Psi _{c}^{\left( +\right) }\left(
\boldsymbol{r}\right) \rangle $ with respect to the incident state $|\Psi
_{c}^{\left( 0\right) }\left( \boldsymbol{r}\right) \rangle $. The
scattering state can be obtained with the same method as in Sec. II. As
shown in appendix A and Ref.~\cite{petrov01}, in the region of $r\gtrsim
r_{\ast }$, the scattering state wave function $|\Psi _{c}^{\left( +\right)
}\left( \boldsymbol{r}\right) \rangle $
\begin{equation}
|\Psi _{c}^{\left( +\right) }\left( \boldsymbol{r}\right) \rangle \approx
|\Psi _{c}^{\left( 0\right) }\left( \boldsymbol{r}\right) \rangle +\frac{A_{%
\mathrm{eff}}\left( c\right) }{\varphi _{0}\left( 0\right) }G(\varepsilon
_{c};\boldsymbol{r},\boldsymbol{0})|0,0\rangle ,  \label{3}
\end{equation}%
with the quasi-2D free Green's function $G\left( \eta ;\boldsymbol{\rho },%
\boldsymbol{\rho }^{\prime }\right) $ given by
\begin{equation}
G\left( \eta ;\boldsymbol{r},\boldsymbol{r}^{\prime }\right) =\frac{1}{\eta
+i0^{+}\!-\!\left[ H_{0}^{\left( \mathrm{2D}\right) }+H_{z}\right] \!}\delta
\left( \boldsymbol{r}-\boldsymbol{r}^{\prime }\right) .
\end{equation}%
Here $A_{\mathrm{eff}}\left( c\right) $ is a $\boldsymbol{r}$-independent
coefficient, and can be derived with the following two facts. First, it can
be proved (see Appendix B) that in the small-distance region $r_{\ast
}\lesssim r\ll 1/k$ the function $|\Psi _{c}^{\left( +\right) }\left(
\boldsymbol{r}\right) \rangle $ behaves as
\begin{equation}
\langle 0,0|\Psi _{c}^{\left( +\right) }\left( \boldsymbol{r}\right) \rangle
\propto \left( \frac{1}{r}-\frac{1}{a}\right)  \label{1}
\end{equation}%
with $a$ the $s$-wave scattering length determined by $\xi $ and the detail
of $V_{\mathrm{3D}}\left( r\right) $ (Appendix B). Second, Appendix C also
shows that, in the small-distance region we have
\begin{eqnarray}
\ \langle 0,0|G(\eta ;\boldsymbol{r},\boldsymbol{0})|0,0\rangle \approx -%
\frac{1}{4\pi r}-\frac{w\left( \frac{\varepsilon _{c}}{2\omega }\right) }{%
2\left( 2\pi \right) ^{3/2}l_{0}} &&  \notag \\
+\sum_{n_{z}=0}^{\infty }\!\left\vert \varphi _{n_{z}}\!\!\left( 0\right)
\right\vert ^{2}\!\!\lambda (\varepsilon _{c}-n_{z}\omega ;\boldsymbol{q}),
&&  \label{2}
\end{eqnarray}%
where $l_{0}=\sqrt{1/\omega }$, $w\left( \eta \right) $ is defined as
\begin{eqnarray}
&&w\left( \eta \right) =  \notag \\
&&\lim_{N\rightarrow \infty }\left( 2\sqrt{\frac{N}{\pi }}\ln \frac{N}{e^{2}}%
-\sum_{j=0}^{N}\frac{\left( 2j-1\right) !!}{\left( 2j\right) !!}\ln \left(
j-\eta -i0^{+}\right) \right)  \notag \\
&&  \label{w}
\end{eqnarray}%
and $\lambda (\eta ;\boldsymbol{q})$ is defined in Eq.~(\ref{lam2d}).
Substituting Eqs.~(\ref{3}) and (\ref{2}) into (\ref{1}), we can obtain the
parameter $A_{\mathrm{eff}}\left( c\right) $. We find that $A_{\mathrm{eff}%
}\left( c\right) $ can be formally expressed as in Eq.~(\ref{a2d}), with the
2D effective range $d$ replaced by a quasi-2D effective range $d_{\mathrm{eff%
}}$
\begin{eqnarray}
&&A_{\mathrm{eff}}\left( c\right) =  \notag \\
&&\frac{\left( 2\pi \right) \langle 0,0|\psi _{c}^{\left( 0\right) }\left(
\boldsymbol{0}\right) \rangle }{i\pi /2-\!C-\!\ln \left\{ d_{\mathrm{eff}%
}\left( \varepsilon _{c},\boldsymbol{q}\right) \sqrt{\varepsilon _{c}}%
/2\right\} -\!\left( 2\pi \right) \lambda \left( \varepsilon _{c},%
\boldsymbol{q}\right) }.  \notag \\
&&  \label{aeff}
\end{eqnarray}%
Here, the effective characteristic length $d_{\mathrm{eff}}$ is a function
of the scattering energy $\varepsilon _{c}$, the center-of-mass momentum $%
\boldsymbol{q}$, and the 3D scattering length $a$, and can be determined by
\begin{eqnarray}
&&\ln d_{\mathrm{eff}}\left( \varepsilon _{c},\boldsymbol{q}\right) =-\frac{%
\sqrt{2\pi }w\left( \frac{\varepsilon _{c}}{2\omega }\right) }{2}-\ln \left(
-\frac{i\sqrt{\varepsilon _{c}}}{2}\right) -C  \notag \\
&&-\frac{\pi l_{0}}{a}+\left( 2\pi \right) ^{2}l_{0}\sum_{n_{z}=1}^{\infty
}\!\left\vert \varphi _{n_{z}}\left( 0\right) \right\vert ^{2}\!\lambda
(\varepsilon _{c}-n_{z}\omega ;\boldsymbol{q}).  \label{deff}
\end{eqnarray}%
Note that the summation of $n_{z}$ on the right-hand-side of Eq.~(\ref{deff}%
) runs over all natural numbers. One can easily show that $\ln d_{\mathrm{eff%
}}\left( \eta ,\boldsymbol{q}\right) $ always takes a real value when $\eta
<\varepsilon _{\mathrm{thre}}({\boldsymbol{q}})+\omega $. We would like to
emphasize that the term $\ln d_{\mathrm{eff}}$ governs the effective 2D
physics in this quasi-2D system, and includes all effects given by the
control parameters $a$ and $\omega $.

\subsection{Effective 2D scattering amplitude}

The straightforward calculation shows that, in the region $\rho \gg l_{0}$
we have%
\begin{eqnarray}
&&|\Psi _{c}^{\left( +\right) }\left( \boldsymbol{r}\right) \rangle
\!\approx \!\varphi _{0}\left( z\right) \left( |\psi _{c}^{\left( 0\right)
}\left( \boldsymbol{\rho }\right) \rangle \!+\!A_{\mathrm{eff}}\left(
c\right) g\left( \varepsilon _{c};\boldsymbol{\rho },\boldsymbol{0}\right)
|0,0\rangle \!\right) ,  \notag \\
&&  \label{psiq2d2}
\end{eqnarray}%
where the 2D Green's function $g$ is defined in Eq.~(\ref{g2d}). Comparing
Eq.~(\ref{psiq2d2}) and Eq.~(\ref{sw2}), we find that in the long-range
region with $\rho \gg l_{0}$, the quasi-2D scattering state wave function $%
|\Psi _{c}^{\left( +\right) }\left( \boldsymbol{r}\right) \rangle \!\!$ is
the product of $\varphi _{0}\left( z\right) $ and a 2D scattering-state wave
function. Therefore, if we focus on the long-range region, the quasi-2D
scattering process is equivalent to a 2D scattering process with an
effective characteristic length $d_{\mathrm{eff}}\left( \varepsilon _{c},%
\boldsymbol{q}\right) $, which is controlled by the 3D scattering length $a$
and the trapping frequency $\omega $ in the $z$-direction.

\begin{figure}[tbp]
\includegraphics[bb=35bp 211bp 566bp 642bp,clip,width=9cm]{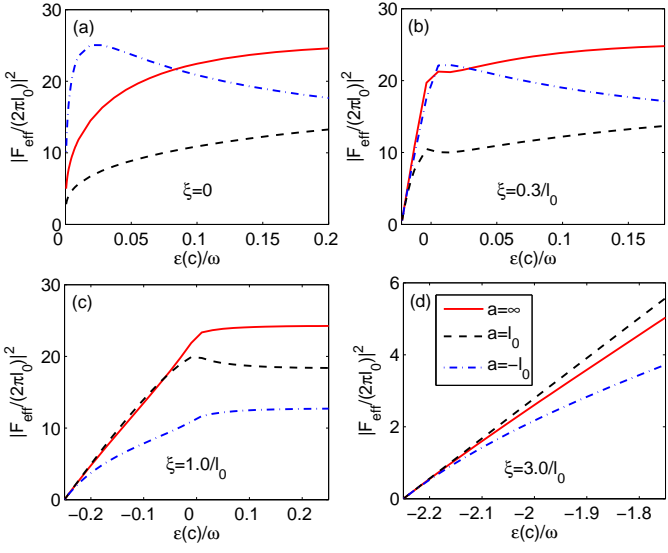}
\caption{(Color online) The variation of $|F_{\mathrm{eff}}|^{2}$ as
functions of the quasi-2D scattering energy $\protect\varepsilon (c)$. Here,
$F_{\mathrm{eff}}$ is defined in Eq.~(\protect\ref{feffa}) and $l_{0}=1/%
\protect\sqrt{\protect\omega }$. We show the results with zero
center-of-mass momentum $q=0$, SOC intensities $\protect\xi =$ (a) $0$, (b) $%
0.3/l_{0}$, (c) $1/l_{0}$, (d) $3/l_{0}$ and 3D scattering lengths $a=\infty
$ (red solid line), $l_{0}$ (black dashed line) and $-l_{0}$ (blue
dashed-dotted line).}
\end{figure}

The quasi-2D scattering amplitude between the incident state $|\Psi
_{c}^{\left( 0\right) }\left( \boldsymbol{r}\right) \rangle $ and an
energy-conserved output state $|\Psi _{c^{\prime }}^{\left( 0\right) }\left(
\boldsymbol{r}\right) \rangle $ with $\boldsymbol{q}=\boldsymbol{q}^{\prime
} $ and $\varepsilon _{c^{\prime }}=\varepsilon _{c}$ is defined as%
\begin{equation}
f^{\left( \mathrm{2D}\right) }\left( c^{\prime }\leftarrow c\right) =\!-2\pi
^{2}\!\!\int \!\!d\boldsymbol{r}^{\prime }\langle \Psi _{c^{\prime
}}^{(0)}\left( \boldsymbol{r}^{\prime }\right) |\ V_{\mathrm{3D}}\!\left(
r^{\prime }\right) |\Psi _{c}^{(+)}\left( \boldsymbol{r}^{\prime }\right)
\rangle .
\end{equation}%
As shown in appendix D, $f^{\left( \mathrm{Q2D}\right) }\left( c^{\prime
}\leftarrow c\right) $ can be expressed as
\begin{equation}
f^{\left( \mathrm{Q2D}\right) }\left( c^{\prime }\leftarrow c\right)
=\!-2\pi ^{2}\!\langle \psi _{c^{\prime }}^{\left( 0\right) }(\boldsymbol{0}%
)|0,0\rangle A_{\mathrm{eff}}\left( c\right) .  \label{feff}
\end{equation}%
Comparing Eq.~(\ref{feff}) with (\ref{t2}), we find that $f^{\mathrm{\left(
Q2D\right) }}$ is nothing but the 2D scattering amplitude with respect to
the 2D scattering state given by the right-hand side of Eq.~(\ref{psiq2d2}).

To understand the behavior of the effective quasi-2D scattering amplitude $%
f^{\mathrm{(Q2D)}}$, we show in Figs.~2 and~3 the variation of the mode
square of the function
\begin{equation}
F_{\mathrm{eff}}\equiv \frac{f^{\left( \mathrm{Q2D}\right) }\left( c^{\prime
}\leftarrow c\right) }{\langle \psi _{c^{\prime }}^{\left( 0\right) }(%
\boldsymbol{0})|0,0\rangle \langle 0,0|\psi _{c}^{\left( 0\right) }(%
\boldsymbol{0})\rangle }  \label{feffa}
\end{equation}%
with scattering energy $\varepsilon _{c}$, SOC intensity $\xi $, 3D
scattering length $a$ and characteristic length $l_{0}$ of the $z$%
-confinement. In Fig. 2, it is shown clearly that the quantity $\left\vert
F_{\mathrm{eff}}\right\vert ^{2}$ linearly decays to zero in the low-energy
limit $\varepsilon _{c}\rightarrow \varepsilon _{\mathrm{thre}}(q)$ with
SOC, and logarithmically decays to zero when there is no SOC. This
observation is consistent with the outcome in the pure 2D systems as
discussed in the previous section. In Fig. 3 we investigate the behavior of $%
\left\vert F_{\mathrm{eff}}\right\vert ^{2}$ as functions of the scattering
length for different values of $\varepsilon _{c}-\varepsilon _{\mathrm{thre}%
}(q)$. Note that for a given value of $\varepsilon _{c}-\varepsilon _{%
\mathrm{thre}}(q)$, the resonance behavior of $\left\vert F_{\mathrm{eff}%
}\right\vert ^{2}$ is still maintained, while the resonance point is shifted
by the SOC and the amplitude of $\left\vert F_{\mathrm{eff}}\right\vert ^{2}$
is suppressed by the SOC.

\begin{figure}[tbp]
\includegraphics[bb=45bp 203bp 581bp 634bp,clip,width=9cm]{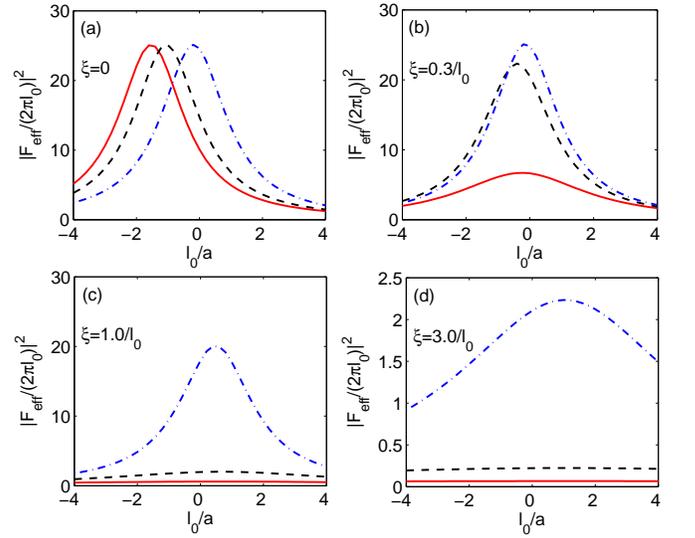}
\caption{(Color online) The variation of $|F_{\mathrm{eff}}|^2$ with 3D
scattering length. The function $F_{\mathrm{eff}}$ is defined in Eq.~(%
\protect\ref{feffa}) and $l_0=1/\protect\sqrt{\protect\omega}$. We show the
results with zero center-of-mass momentum $q=0$, SOC intensities $\protect\xi%
=$ (a) $0$, (b) $0.3/l_0$, (d) $1/l_0$, (d) $3/l_0$ and the quasi-2D
scattering energy $\protect\varepsilon(c)=-\protect\xi^2/4+0.006\protect%
\omega$ (red solid line), $-\protect\xi^2/4+0.02\protect\omega$ (black
dashed line) and $-\protect\xi^2/4+0.2\protect\omega$ (blue dashed-dotted
line).}
\end{figure}

\subsection{Two-atom bound state}

Next, we consider the two-body bound state in quasi-2D configuration.
Similar as in Sec. II, it can be shown that in the region $r\gtrsim r_{\ast
} $ the wave function $|\Psi _{b}\left( \boldsymbol{r}\right) \rangle $ can
be expressed as
\begin{equation}
|\Psi _{b}\left( \boldsymbol{r}\right) \rangle =B^{\prime }G\left( E_{b};%
\boldsymbol{r},\boldsymbol{0}\right) |0,0\rangle  \label{tab}
\end{equation}%
with $B^{\prime }$ the normalization factor. In the long-range limit $r\gg
l_{0}$, the wave function of the quasi-2D bound state is proportional to
that of the pure-2D bound state in Eq.~(\ref{psib2d}). The energy $E_{b}$ of
the quasi-2D bound state is determined by the boundary condition in the
small-distance region (appendix B)
\begin{equation}
\langle 0,0|\Psi _{b}\left( \boldsymbol{r}\right) \rangle \propto \frac{1}{r}%
-\frac{1}{a}  \label{r}
\end{equation}%
Then it is easy to prove that the boundary condition Eq.~(\ref{r}) is
equivalent to the equation
\begin{equation}
-\ln d_{\mathrm{eff}}\left( E_{b},\boldsymbol{q}\right) =C+\ln \left( -\frac{%
i\sqrt{E_{b}}}{2}\right) +2\pi \lambda (E_{b},\boldsymbol{q}).  \label{ebq2d}
\end{equation}

\begin{figure}[tbp]
\includegraphics[bb=47bp 323bp 404bp 634bp,clip,width=7cm]{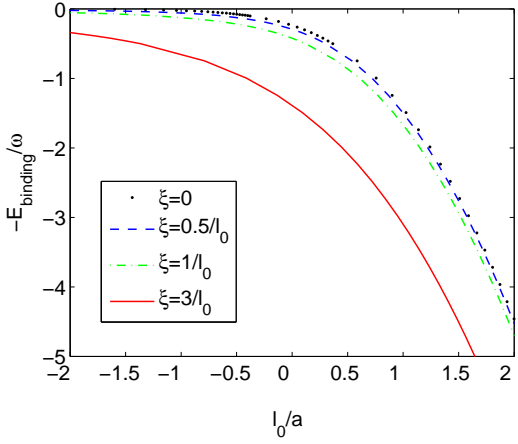}
\caption{(Color online) The binding energy $E_{\mathrm{binding}}= \protect%
\varepsilon_{\mathrm{thre}}(q)-E_{b}$ of the quasi-2D two-atom bound state
as a function of the 3D scattering length. In this plot, we set the
center-of-mass momentum $q=0$ and the SOC intensities $\protect\xi=0$ (black
dots), $0.5/l_0$ (blue dashed line), $1/l_0$ (green dashed-dotted line) and $%
3/l_0$ (red solid line).}
\end{figure}

In Fig. 4, we show the behavior of binding energy
\begin{equation}
E_{\mathrm{binding}}=\varepsilon _{\mathrm{thre}}(q)-E_{b}
\end{equation}
with varying 3D scattering length $a$ and SOC intensity for cases of $q=0$.
Notice that the variation of $E_{\mathrm{binding}}$ in terms of $1/a$ has
the same qualitative behavior with or without SOC, and the value of $E_{%
\mathrm{binding}}$ is increased with the SOC intensity $\xi$.

Next, we discuss the dispersion relation of the two-atom bound state. To
this end, we express $E_{b}$ as $E_{b}=\sum_{n=0}^{\infty
}E_{bn}q^{n}h_{n}\left( \boldsymbol{q}/q\right) $ and then substitute this
expression into Eq.~(\ref{ebq2d}). Expanding both sides of Eq.~(\ref{ebq2d}%
), we can obtain all the coefficients $E_{bn}$ and the functions $h_{n}$.
When $q$ is small, we have
\begin{equation}
E_{b}\approx E_{b0}+E_{b2}q^{2}  \label{ebt}
\end{equation}%
with $E_{b0}$ and $E_{b2}$ determined by the equations%
\begin{eqnarray}
\frac{l_{0}\sqrt{2\pi }}{2a} &=&\left( 2\pi \right)
^{3/2}l_{0}\sum_{n_{z}=0}^{\infty }\!\left\vert \varphi _{n_{z}}\left(
0\right) \right\vert ^{2}\!\lambda _{0}(E_{b0}-n_{z}\omega )  \notag \\
&&-\frac{1}{2}w\left( \frac{E_{b0}}{2\omega }\right) ;
\end{eqnarray}%
and
\begin{eqnarray}
E_{b2}\! &=&\!\frac{\sum_{n_{z}=0}^{\infty }\!\left\vert \varphi
_{n_{z}}\left( 0\right) \right\vert ^{2}\!\lambda _{2}(E_{b0}-n_{z}\omega )}{%
\frac{1}{4(2\pi )^{3/2}\omega l_{0}}w^{\prime }\!\left( \frac{E_{b0}}{%
2\omega }\right) \!\!-\!\!\sum_{n_{z}=0}^{\infty }\!\left\vert \varphi
_{n_{z}}\!\!\left( 0\right) \right\vert ^{2}\!\lambda _{0}^{\prime
}(E_{b0}\!-\!n_{z}\omega )}.  \notag \\
&&
\end{eqnarray}%
Here, the functions $\lambda _{0}\left( \eta \right) $ and $\lambda
_{2}(\eta )$ are defined as
\begin{equation}
\lambda _{0}\left( \eta \right) =-\frac{\xi \left( \pi +2\arctan \left[
\frac{\xi ^{2}+2\eta }{\xi \sqrt{-4\eta -\xi ^{2}}}\right] \right) }{8\pi
\sqrt{-4\eta -\xi ^{2}}}
\end{equation}%
and%
\begin{eqnarray}
&&\lambda _{2}\left( \eta \right)  \notag \\
&=&\frac{\xi \left( \xi \sqrt{-4\eta -\xi ^{2}}+2(2\eta +\xi ^{2})\arctan %
\left[ \frac{\xi }{\sqrt{-4\eta -\xi ^{2}}}\right] \right) }{32\pi \eta
\left( -4\eta -\xi ^{2}\right) ^{3/2}}  \notag \\
&&
\end{eqnarray}%
with $w^{\prime }\left( \eta \right) =dw\left( \eta \right) /d\eta $ and $%
\lambda _{0}^{\prime }(\eta )=d\lambda _{0}\left( \eta \right) /d\eta $.

The total energy of the two-body bound state can then be obtained by adding
Eq.~(\ref{ebt}) and the kinetic energy of the center-of-mass motion, leading
to $E_{bt}\approx E_{b0}+E_{b2}q^{2}+q^{2}/4$. This quantity can also be
expressed in terms of the effective mass
\begin{equation}
m_{\mathrm{eff}}=\frac{1}{2E_{b2}+1/2},  \label{meff}
\end{equation}%
and takes the form
\begin{equation}
E_{b2}\approx E_{b0}+\frac{q^{2}}{2m_{\mathrm{eff}}}.
\end{equation}%
In Fig. 5, we plot the effective mass with the 3D scattering length and the
SOC intensity.

\begin{figure}[tbp]
\includegraphics[bb=45bp 406bp 558bp 634bp,clip,width=8.5cm]{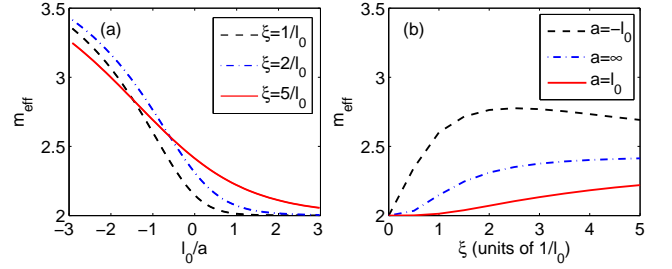}
\caption{(Color online) (a) The variation of the effective mass $m_{\mathrm{%
eff}}$ of the two-atom bound state in the quasi-2D system as a function of
the 3D scattering length. The SOC intensities used in this plot are $\protect%
\xi =1/l_{0}$ (black dashed line), $2/l_{0}$ (blue dashed-dotted line) and $%
5/l_{0}$ (red solid line). (b) The variation of $m_{\mathrm{eff}}$ as a
function of the SOC intensity $\protect\xi $. The 3D scattering lengths used
in this plot are $a=-l_{0}$ (black dashed line), $\infty $ (blue
dashed-dotted line) and $l_{0}$ (red solid line).}
\end{figure}


\section{Effective 2D Hamiltonian}

With the above knowledge of two-body physics, we can construct the effective
2D Hamiltonian for a two-component Fermi gas with Rashba SOC in a quasi-2D
confinement. The effective Hamiltonian is required to give the same
scattering amplitude or two-body bound state as the original Hamiltonian. It
is pointed out that, for each effective 2D Hamiltonian, this criteria can
only be satisfied for a small energy range. Thus, in principle one needs to
derive different effective Hamiltonian for different energy regions.

We first consider gases with atoms in the low-energy scattering states i.e.,
the system given by the directly cooling of the atoms with a fixed
scattering length. In such a system, the energy of relative motion of two
atoms is slightly above $-\xi ^{2}/4$, and the probability for atoms in the
excited states along the $z$-direction is negligible. The atom-atom
interaction can be described by a 2D single-channel contact potential%
\begin{equation}
\hat{V}_{o}=\frac{1}{\mathcal{S}}\sum_{\boldsymbol{k},\boldsymbol{k}^{\prime
},\boldsymbol{k^{\prime\prime}}}\,^{\prime }g(\boldsymbol{k}+\boldsymbol{k}%
^{\prime })a_{\boldsymbol{k},\uparrow }^{\dagger }a_{\boldsymbol{k}^{\prime
},\downarrow }^{\dagger }a_{\boldsymbol{k}^{\prime }+\boldsymbol{%
k^{\prime\prime}},\downarrow }a_{\boldsymbol{k}^{\prime }-\boldsymbol{%
k^{\prime\prime}},\uparrow }.  \label{vo}
\end{equation}%
Here, $a_{\boldsymbol{p},\sigma }^{\dagger }$ and $a_{\boldsymbol{p},\sigma
} $ are the 2D creation and annihilation operators for an atom with momentum
$\boldsymbol{p}$ and spin $\sigma $, $\mathcal{S}$ is the area of the
system, and the summation $\sum_{\boldsymbol{k},\boldsymbol{k}^{\prime },%
\boldsymbol{k^{\prime\prime}}}^{\prime }$ is done for $|\boldsymbol{k}-%
\boldsymbol{k}^{\prime }|/2<k_{c} $ and $|(\boldsymbol{k}-\boldsymbol{k}%
^{\prime })/2+\boldsymbol{k^{\prime\prime}}|<k_{c}$, with $k_{c}$ a cut-off
momentum. For systems without SOC, the coupling intensity $g(%
\boldsymbol{q})$ is given by the renormalization relation~\cite{without}
$1/(2\pi g)=-\int_{k<k_{c}}d\boldsymbol{k}(E_{b}+k^{2})^{-1}$, with $E_{b}$
the bound-state energy. For our current systems with Rashba SOC, the
renormalization relation should be modified as%
\begin{equation}
\ln k_{c}-\frac{1}{2\pi g(\boldsymbol{q})}=-C-\ln \left[ \frac{d_{\mathrm{eff%
}}\left( -\frac{\xi ^{2}}{4},\boldsymbol{q}\right) }{2}\right]  \label{re1}
\end{equation}%
with the effective characteristic length $d_{\mathrm{eff}}$ defined in Eq.~(%
\ref{deff}). According to the definition, $d_{\mathrm{eff}}$ also depends on
3D scattering length $a$ and the characteristic length $l_{0}$ of the $z$
confinement. A straightforward calculation (Appendix F) shows that the 2D
scattering amplitude given by the interaction potential $V_{o}$ is the same
as the quasi-2D scattering amplitude $f^{\left( \mathrm{Q2D}\right) }$ as in
Eq.~(\ref{feff}).

Now we consider the systems with atoms in the bound states. In that case, if
the binding energy of the bound state is large, the atomic population in the
transverse excited states $\varphi _{n_{z}}(z)$ with $n_{z}>0$ becomes
significant~\cite{Duan-07,Duan-06}. To take into account these transverse
excitations, we use a phenomenological two-channel model to describe the
atom-atom interaction:%
\begin{eqnarray}
\hat{V}_{b} &=&\hat{V}_{o}+\sum_{\boldsymbol{q}}\left( \frac{q^{2}}{4}+v(%
\boldsymbol{q})\right) b_{\boldsymbol{q}}^{\dagger }b_{\boldsymbol{q}}
\notag \\
&&+\frac{1}{\sqrt{\mathcal{S}}}\sum_{\boldsymbol{k},\boldsymbol{k}^{\prime
}}\,^{\prime }u(\boldsymbol{k}+\boldsymbol{k}^{\prime })a_{\boldsymbol{k}%
,\uparrow }^{\dagger }a_{\boldsymbol{k}^{\prime },\downarrow }^{\dagger }b_{%
\boldsymbol{k+k}^{\prime }}+h.c.
\end{eqnarray}%
Here, the open-channel interaction $\hat{V}_{o}$ is defined as in Eq.~(\ref%
{vo}), $b_{\boldsymbol{q}}^{\dagger }$ and $b_{\boldsymbol{q}}$ are the
creation and annihilation operators for molecules in the closed channel of
our phenomenological model with $[b_{\boldsymbol{q}},b_{\boldsymbol{q}%
^{\prime }}^{\dagger }]=\delta _{\boldsymbol{q},\boldsymbol{q}^{\prime }}$,
and the summation $\sum_{\boldsymbol{k},\boldsymbol{k}^{\prime }}^{\prime }$
is done for $|\boldsymbol{k}-\boldsymbol{k}^{\prime }|/2<k_{c}$. The
renormalization relation for cases without SOC is given in
Ref.~\cite{Duan-06}. In the presence of Rashba SOC, The parameters $%
v(\boldsymbol{q})$ and $u(\boldsymbol{q})$ are now given by
\begin{equation}
\frac{1}{\kappa \left( E_{b},\boldsymbol{q}\right) }=g(\boldsymbol{q})+\frac{%
u(\boldsymbol{q})^{2}}{E_{b}-v(\boldsymbol{q})^{2}}  \label{re2}
\end{equation}%
and%
\begin{equation}
\frac{\left( \frac{u(\boldsymbol{q})}{E_{b}-v(\boldsymbol{q})}\right) ^{2}}{%
\left( g(\boldsymbol{q})+\frac{u(\boldsymbol{q})^{2}}{E_{b}-v(\boldsymbol{q})%
}\right) ^{2}} \\
=\frac{\sum_{n_{z}=1}^{\infty }\left\vert \phi _{n_{z}}\left( 0\right)
\right\vert ^{2}\chi \left( E_{b}-n_{z}\omega ,\boldsymbol{q}\right) }{%
\left\vert \phi _{0}\left( 0\right) \right\vert ^{2}},  \label{re3}
\end{equation}%
where the functions $\kappa \left( \eta ,\boldsymbol{q}\right) $ and $\chi
\left( \eta ,\boldsymbol{q}\right) $ are defined as
\begin{equation}
\kappa \left( \eta ,\boldsymbol{q}\right) =-2\pi \ln k_{c}+\pi \ln
\left\vert \eta \right\vert +\left( 2\pi \right) ^{2}\lambda \left( \eta ,%
\boldsymbol{q}\right)
\end{equation}%
and%
\begin{equation}
\chi \left( \eta ,\boldsymbol{q}\right) =\sum_{\boldsymbol{\alpha }%
}\int_{k<k_c}d\boldsymbol{k}\frac{\left\vert \langle 0,0|\boldsymbol{\alpha }%
(\boldsymbol{q},\boldsymbol{k})\rangle \right\vert ^{2}}{\left( \eta
-\varepsilon_c\right) ^{2}%
}\,,
\end{equation}%
with $c=(\boldsymbol{\alpha },\boldsymbol{q},\boldsymbol{k})$. 
In Appendix F, we show that the 2D bound state $|\Psi _{b}^{\left( \mathrm{%
eff}\right) }(\boldsymbol{r})\rangle $ given by $V_{b}$ has the same binding
energy as the quasi-2D bound state $|\Psi _{b}(\boldsymbol{r})\rangle $
obtained in the previous section. Besides, it is also proven that the
open-channel probability of $|\Psi _{b}^{\left( \mathrm{eff}\right) }(%
\boldsymbol{r})\rangle $ recovers the population of the transversal
ground-state $|\Psi _{b}(\boldsymbol{r})\rangle $. Therefore, the effective
2D potential $V_{b}$ can be used to study low-energy physics where nearly
all particles are in the bound states with zero center-of-mass momentum.

\section{Conclusions}

In this paper we investigate the two-body physics of two-component fermonic
atoms in 2D and quasi-2D configurations with a Rashba SOC. For the 2D case,
we find that when the total momentum is zero, the logarithmic behavior of
the 2D scattering amplitude in the low-energy limit is replaced by a
polynomial behavior. For the quasi-2D system, we obtain an analytic
expression of the effective 2D scattering amplitude as a function of the 3D
scattering length $a$ and the trapping frequency $\omega$ along the strongly
confined $z$-direction, and observe the same polynomial behavior. We also
discuss the two-atom bound state, and get the algebraic equation for the
binding energy. We find that the two-body binding energy is enhanced by the
presence of SOC, as a consequence of the increase of single particle density
of states in the low energy limit. We also analyze the dispersion relation
and extract the effective mass of the dimers. These information allows us to
tune the effective 2D physics in quasi-2D configuration with parameters $a$
and $\omega$.

With the knowledge of two-body physics, we further construct two effective
2D Hamiltonians, which can individually mimic the original quasi-2D
Hamiltonian within the energy regimes around the two-body binding energy or
close to the single-particle threshold. These effective models can be used
to analyze many-body physics of the system, hence help paving a way towards
the simulation of 2D Fermi system with quasi-2D atomic gases. Our method
developed in this paper can be directly generalized to systems of bosonic or
distinguishable atoms with arbitrary type of SOC.

\begin{acknowledgments}
We thank Xiaoling Cui and Hui Zhai for helpful discussion. This work is
supported by National Natural Science Foundation of China (11074305,
10904172), the NKBRSF of China (Grant No. 2012CB922104), the Fundamental
Research Funds for the Central Universities, and the Research Funds of
Renmin University of China (10XNL016). PZ and WZ would also like to thank
the NCET Program for support.
\end{acknowledgments}

\appendix
\addcontentsline{toc}{section}{Appendices}\markboth{APPENDICES}{}
\begin{subappendices}

\section{wave functions of 2D scattering states and bound states}

In this appendix we derive the Eqs.~(\ref{sw2}, \ref{3}) for 2D and quasi-2D
low-energy scattering states, as well as Eqs.~(\ref{psib2d}, \ref{tab}) for
2D and quasi-2D bound states.

We first prove Eq.~(\ref{sw2}) for the 2D scattering wave function $|\psi
_{c}^{(+)}\left( \boldsymbol{\rho }\right) \rangle $. The Lippmann-Schwinger
equation gives%
\begin{equation}
|\psi _{c}^{(+)}\left( \boldsymbol{\rho }\right) \rangle =|\psi
_{c}^{(0)}\left( \boldsymbol{\rho }\right) \rangle +\int d\boldsymbol{\rho }%
^{\prime }g(\varepsilon _{c};\boldsymbol{\rho },\boldsymbol{\rho }^{\prime
})V_{\mathrm{2D}}\!\left( \boldsymbol{\rho }^{\prime }\right) |\psi
_{c}^{(+)}\left( \boldsymbol{\rho }^{\prime }\right) \rangle .  \label{a1}
\end{equation}%
Here, $|\psi _{c}^{(0)}\left( \boldsymbol{\rho }\right) \rangle $ is the
incident state and the free Green's function $g(\varepsilon _{c};\boldsymbol{%
\rho },\boldsymbol{\rho }^{\prime })$ is defined in Eq.~(\ref{g2d}). Since
the potential $V_{\mathrm{2D}}\!\left( \boldsymbol{\rho }\right) $ is
negligible in the region $\rho \gtrsim \rho _{\ast }$, the integration in
Eq.~(\ref{a1}) is only effective in the region $\rho ^{\prime }\lesssim \rho
_{\ast }$. On the other hand, in the low-energy cases $k\ll 1/\rho _{\ast }$%
, when $\rho \rightarrow \infty $ and $\rho ^{\prime }\lesssim \rho _{\ast }$%
, the function $g(\varepsilon _{c};\boldsymbol{\rho },\boldsymbol{\rho }%
^{\prime })$ becomes very steady with respect to $\boldsymbol{\rho }^{\prime
}$ and we have
\begin{equation}
g(\varepsilon _{c};\boldsymbol{\rho },\boldsymbol{\rho }^{\prime })\approx
g(\varepsilon _{c};\boldsymbol{\rho },\boldsymbol{0}).
\end{equation}%
Therefore, in the limit $\rho \rightarrow \infty $, the solution of Eq.~(\ref%
{a1}) takes the form
\begin{equation}
|\psi _{c}^{(+)}\left( \boldsymbol{\rho }\right) \rangle =|\psi
_{c}^{(0)}\left( \boldsymbol{\rho }\right) \rangle +g(\varepsilon _{c};%
\boldsymbol{\rho },\boldsymbol{0})|\chi \rangle ,  \label{sww}
\end{equation}%
where the spin state $|\chi \rangle $ is related to $|\psi _{c}^{(+)}\left(
\boldsymbol{\rho }\right) \rangle $ via the equation
\begin{equation}
|\chi \rangle =\int d\boldsymbol{\rho }^{\prime }V_{\mathrm{2D}}\!\left(
\boldsymbol{\rho }^{\prime }\right) |\psi _{c}^{(+)}\left( \boldsymbol{\rho }%
^{\prime }\right) \rangle .  \label{aa}
\end{equation}

On the other hand, since $|\psi _{c}^{(+)}\left( \boldsymbol{\rho }\right)
\rangle $ is an eigen-state of $H_{0}^{\left( \mathrm{2D}\right) }(%
\boldsymbol{q})+V_{\mathrm{2D}}\!\left( \boldsymbol{\rho }\right) $ and the
potential $V_{\mathrm{2D}}\left( \boldsymbol{\rho }\right) $ is negligible
in the region $\rho \gtrsim \rho _{\ast }$, in such a region the wave
function $|\psi _{c}^{(+)}\left( \boldsymbol{\rho }\right) \rangle $
satisfies the equation
\begin{equation}
H_{0}^{\left( \mathrm{2D}\right) }|\psi _{c}^{(+)}\left( \boldsymbol{\rho }%
\right) \rangle =\varepsilon _{c}|\psi _{c}^{(+)}\left( \boldsymbol{\rho }%
\right) \rangle .  \label{sse}
\end{equation}

Therefore, the behavior of the wave function $|\psi _{c}^{(+)}\left(
\boldsymbol{\rho }\right) \rangle $ in the region $\rho \gtrsim \rho _{\ast
} $ is determined by Eq.~(\ref{sse}) and the boundary condition (\ref{sww})
in the limit $\rho \rightarrow\infty$. It is easy to prove that, the
function $|\psi _{c}^{(0)}\rangle +g(\varepsilon _{c};\boldsymbol{\rho },%
\boldsymbol{0})|\chi \rangle $ satisfies both of the two conditions.
Therefore, $|\psi _{c}^{(+)}\left( \boldsymbol{\rho }\right) \rangle $
satisfies Eq.~(\ref{sse}) not only in the limit $\rho \rightarrow\infty$,
but also in the entire region of $\rho \gtrsim \rho _{\ast }$.

Furthermore, due to the facts $\hat{P}_{12}V_{\mathrm{2D}}\!\left(
\boldsymbol{\rho }\right) \hat{P}_{12}=V_{\mathrm{2D}}\!\left( \boldsymbol{%
\rho }\right) $ and $\hat{P}_{12}|\psi _{c}^{(+)}\left( \boldsymbol{\rho }%
\right) \rangle =-|\psi _{c}^{(+)}\left( \boldsymbol{\rho }\right) \rangle $
with $\hat{P}_{12}$ the permutation operator of the two atoms, the
integration (\ref{aa}) can be re-written as%
\begin{eqnarray}
&&\int d\boldsymbol{\rho }^{\prime }V_{\mathrm{2D}}\!\left( \boldsymbol{\rho
}^{\prime }\right) |\psi _{c}^{(+)}\left( \boldsymbol{\rho }^{\prime
}\right) \rangle  \notag \\
&=&|0,0\rangle \int d\boldsymbol{\rho }^{\prime }V_{\mathrm{2D}}\!\left(
\boldsymbol{\rho }^{\prime }\right) \langle 0,0|\psi _{c}^{(+)}\left(
\boldsymbol{\rho }^{\prime }\right) \rangle .
\end{eqnarray}%
Thus, $|\chi \rangle $ can be expressed as%
\begin{equation}
|\chi \rangle =A\left( c\right) |0,0\rangle  \label{kap}
\end{equation}%
with $\!A\left( c\right) $ a c-number. Substituting Eq.~(\ref{kap}) into (%
\ref{sww}), we finally obtain Eq.~(\ref{sw2}).

It is easy to find that Eqs.~(\ref{3}, \ref{psib2d}) and (\ref{tab}) can be
proved within the same approach. Especially, we have the result
\begin{equation}
\frac{A_{\mathrm{eff}}\left( c\right) }{\varphi _{0}\left( 0\right) }%
|0,0\rangle =\int d\boldsymbol{r}^{\prime }V_{\mathrm{3D}}\!\left( r^{\prime
}\right) |\Psi _{c}^{(+)}\left( \boldsymbol{r}^{\prime }\right) \rangle .
\label{kaka}
\end{equation}%
which is similar as the ones in Eqs. (\ref{aa}) and (\ref{kap}).

\section{small-distance behavior of 2D and quasi-2D wave function}

In this appendix we prove Eqs.~(\ref{src2d}, \ref{1}, \ref{bb1}, \ref{r})
for the behaviors of the 2D and quasi-2D wave functions in the
small-distance region $\rho_{\ast}<<\rho<<1/k$ or $r_{\ast}<<r<<1/k$. Here
we only show the proof of Eqs. (\ref{src2d}, \ref{1}). Eqs. (\ref{bb1}, \ref%
{r}) can be derived with the same method.

We first prove Eq. (\ref{src2d}) for the behavior of $|\psi _{c}^{(+)}\left(
\boldsymbol{\rho }\right) \rangle $. To this end, we define a rotated wave
function%
\begin{equation}
|\tilde{\psi}_{c}^{(+)}\left( \boldsymbol{\rho }\right) \rangle =U\left(
\boldsymbol{\rho }\right) |\psi _{c}^{(+)}\left( \boldsymbol{\rho }\right)
\rangle
\end{equation}%
with the unitary transformation $U\left( \boldsymbol{\rho }\right) $ defined
as%
\begin{equation}
U\left( \boldsymbol{\rho }\right) =\exp \left[ -\frac{i\xi x}{4}\left(
\sigma _{x}^{\left( 1\right) }-\sigma _{x}^{\left( 2\right) }\right) \right]
\exp \left[ -\frac{i\xi y}{4}\left( \sigma _{y}^{\left( 1\right) }-\sigma
_{y}^{\left( 2\right) }\right) \right] .  \label{uuu}
\end{equation}%
Then the eigen-equation%
\begin{equation}
H^{\left( \mathrm{2D}\right) }|\psi _{c}^{(+)}\left( \boldsymbol{\rho }%
\right) \rangle =\varepsilon _{c}|\psi _{c}^{(+)}\left( \boldsymbol{\rho }%
\right) \rangle
\end{equation}%
satisfied by $|\psi _{c}^{(+)}\left( \boldsymbol{\rho }\right) \rangle $
gives%
\begin{equation}
\tilde{H}^{\left( \mathrm{2D}\right) }|\tilde{\psi}_{c}^{(+)}\left(
\boldsymbol{\rho }\right) \rangle =\varepsilon _{c}|\tilde{\psi}%
_{c}^{(+)}\left( \boldsymbol{\rho }\right) \rangle ,
\end{equation}%
where
\begin{eqnarray}
&&\tilde{H}^{\left( \mathrm{2D}\right) }=U\left( \boldsymbol{\rho }\right)
H^{\left( \mathrm{2D}\right) }U^{\dagger }\left( \boldsymbol{\rho }\right)
\notag \\
&=&-\sum_{\beta =x,y}\frac{\partial ^{2}}{\partial \beta ^{2}}-iW\left(
\boldsymbol{\rho }\right) \left( \frac{\partial }{\partial y}\right)
+W^{\prime }\left( \boldsymbol{q},\boldsymbol{\rho }\right) +\tilde{V}_{%
\mathrm{2D}}\left( \rho \right)  \notag \\
&\equiv &\tilde{H}_{0}^{\left( \mathrm{2D}\right) }+\tilde{V}_{\mathrm{2D}%
}\left( \rho \right)  \label{app3}
\end{eqnarray}%
Here, we have $\tilde{V}_{\mathrm{2D}}=UV_{\mathrm{2D}}U^{\dagger }$ and the
operators $W\left( \boldsymbol{\rho }\right) $ and $W^{\prime }\left(
\boldsymbol{q},\boldsymbol{\rho }\right) $ are defined as
\begin{equation}
W\left( \boldsymbol{\rho }\right) =-\xi \left( \sigma _{y}^{\left( 1\right)
}-\sigma _{y}^{\left( 2\right) }\right) +\xi U^{\dagger }\left( \sigma
_{y}^{\left( 1\right) }-\sigma _{y}^{\left( 2\right) }\right) U  \label{www}
\end{equation}%
and
\begin{eqnarray}
W^{\prime }\left( \boldsymbol{q},\boldsymbol{\rho }\right) &=&-\frac{\xi ^{2}%
}{16}\sum_{\alpha =x,y}U^{\dagger }\left( \boldsymbol{\rho }\right) \left(
\sigma _{\alpha }^{\left( 1\right) }-\sigma _{\alpha }^{\left( 2\right)
}\right) ^{2}U\left( \boldsymbol{\rho }\right)  \notag \\
&&+\frac{\xi }{4}\sum_{\alpha =x,y}q_{\alpha }U^{\dagger }\left( \boldsymbol{%
\rho }\right) \left( \sigma _{\alpha }^{\left( 1\right) }-\sigma _{\alpha
}^{\left( 2\right) }\right) U\left( \boldsymbol{\rho }\right)  \notag \\
&&+\frac{W\left( \boldsymbol{\rho }\right) ^{2}}{16}-\frac{i}{2}\frac{%
\partial }{\partial y}W\left( \boldsymbol{\rho }\right) .  \label{wp}
\end{eqnarray}%
Due to the weak SOC condition $\xi <<1/\rho _{\ast }$, in the region $\rho
\ll 4/\xi $ we have we have $W\left( \boldsymbol{\rho }\right) \approx 0$
and $W^{\prime }\left( \boldsymbol{q},\boldsymbol{\rho }\right) \approx
W^{\prime }\left( \boldsymbol{q},\boldsymbol{0}\right) $. Therefore, in this
region the rotated wave function $|\tilde{\psi}_{c}^{(+)}\left( \boldsymbol{%
\rho }\right) \rangle $ is determined by the Hamiltonian
\begin{equation}
-\sum_{\beta =x,y}\frac{\partial ^{2}}{\partial \beta ^{2}}+W^{\prime
}\left( \boldsymbol{q},\boldsymbol{0}\right) +\tilde{V}_{\mathrm{2D}}\left(
\rho \right)
\end{equation}%
without SOC. Then the behavior of $|\tilde{\psi}_{c}^{(+)}\left( \boldsymbol{%
\rho }\right) \rangle $ in the small-distance region is the same as the one
for the 2D scattering wave function between two fermions without SOC, and
can be described as%
\begin{equation}
|\tilde{\psi}_{c}^{(+)}\left( \boldsymbol{\rho }\right) \rangle \propto
\left( \ln \rho -\ln d\right) |0,0\rangle .  \label{ap1}
\end{equation}%
with the characteristic length $d$ determined by the detail of $\tilde{V}_{%
\mathrm{2D}}(\boldsymbol{\rho })$. Using the relation $|\psi
_{c}^{(+)}\left( \boldsymbol{\rho }\right) \rangle =U^{\dag }\left(
\boldsymbol{\rho }\right) |\tilde{\psi}_{c}^{(+)}\left( \boldsymbol{\rho }%
\right) \rangle $, we immediately get the result in Eq. (\ref{src2d}) for
the small-distance behavior of $|\psi _{c}^{(+)}\left( \boldsymbol{\rho }%
\right) \rangle $.

Eq. (\ref{1}) can be proved with the similar approach. We can define the
quasi-2D rotated wave function
\begin{equation}
|\tilde{\Psi}_{c}^{(+)}\left( \boldsymbol{r}\right) \rangle =U\left(
\boldsymbol{r}\right) |\Psi _{c}^{(+)}\left( \boldsymbol{r}\right) \rangle .
\end{equation}%
Then $|\tilde{\Psi}_{c}^{(+)}\left( \boldsymbol{r}\right) \rangle $ satisfies%
\begin{equation}
\left[ \tilde{H}_{0}^{\left( \mathrm{2D}\right) }+H_{z}+\tilde{V}_{\mathrm{3D%
}}\left( \boldsymbol{r}\right) \right] |\tilde{\Psi}_{c}^{(+)}\left(
\boldsymbol{r}\right) \rangle =\varepsilon _{c}|\tilde{\Psi}_{c}^{(+)}\left(
\boldsymbol{r}\right) \rangle ,
\end{equation}%
with $\tilde{H}_{0}^{\left( \mathrm{2D}\right) }$ defined in Eq. (\ref{app3}%
) and $\tilde{V}_{\mathrm{3D}}=UV_{\mathrm{3D}}U^{\dagger }$. Similar as
above, in the region $r\ll 4/\xi $ we have $W\left( \boldsymbol{\rho }%
\right) \approx 0$ and $W^{\prime }\left( \boldsymbol{q},\boldsymbol{\rho }%
\right) \approx W^{\prime }\left( \boldsymbol{q},\boldsymbol{0}\right) $,
and $|\tilde{\Psi}_{c}^{(+)}\left( \boldsymbol{r}\right) \rangle $ is
determined by the Hamiltonian
\begin{equation}
-\sum_{\beta =x,y}\frac{\partial ^{2}}{\partial \beta ^{2}}+H_{z}+W^{\prime
}\left( \boldsymbol{q},\boldsymbol{0}\right) +\tilde{V}_{\mathrm{3D}}\left(
\boldsymbol{r} \right)
\end{equation}%
without SOC. Then the behavior of $|\tilde{\Psi}_{c}^{(+)}\left( \boldsymbol{%
r}\right) \rangle $ in the small-distance region is the same as the one for
the 3D scattering wave function between two fermions without SOC, and can be
described as%
\begin{equation}
|\tilde{\Psi}_{c}^{(+)}\left( \boldsymbol{r}\right) \rangle \propto \left(
\frac{1}{r}-\frac{1}{a}\right) |0,0\rangle ,  \label{ap2}
\end{equation}%
with the scattering length $a$ determined by the detail of $\tilde{V}_{%
\mathrm{3D}}\left( \boldsymbol{r}\right) $. Using the relation $|\Psi
_{c}^{(+)}\left( \boldsymbol{r}\right) \rangle =U^{\dag }\left( \boldsymbol{r%
}\right) |\tilde{\Psi}_{c}^{(+)}\left( \boldsymbol{r}\right) \rangle $, we
obtain the behavior of $|\Psi _{c}^{(+)}\left( \boldsymbol{r}\right) \rangle
$ in the small-distance region:
\begin{equation}
|\Psi _{c}^{(+)}\left( \boldsymbol{r}\right) \rangle \propto \left( \frac{1}{%
r}-\frac{1}{a}\right) |0,0\rangle +\frac{i\xi }{4}\sum_{\beta =x,y}\frac{%
\beta }{r}\left( \sigma _{\beta }^{\left( 1\right) }-\sigma _{\beta
}^{\left( 2\right) }\right) |0,0\rangle .
\end{equation}%
Then we have the result in Eq. (\ref{1}):%
\begin{equation}
\langle 0,0|\Psi _{c}^{(+)}\left( \boldsymbol{r}\right) \rangle \propto
\left( \frac{1}{r}-\frac{1}{a}\right) .
\end{equation}%
In the end of this appendix, we point out that, the results in Eqs. (\ref%
{ap1}) and (\ref{ap2}) of this appendix can be proved more explicitly with
the approach given in Ref. \cite{BP}.

\section{small-distance behavior of 2D and quasi-2D Green's function}

In this appendix we prove Eqs.~(\ref{gg}, \ref{2}) for the behavior of the
functions $\langle 0,0|g(\varepsilon _{c};\boldsymbol{\rho },\boldsymbol{0}%
)|0,0\rangle $ and $\langle 0,0|G(\varepsilon _{c};\boldsymbol{r},%
\boldsymbol{0})|0,0\rangle $ in the small-distance region. We begin from Eq.
(\ref{gg}). In the small-distance region $\rho _{\ast }\lesssim \rho <<1/k$,
the function $\langle 0,0|g(\varepsilon _{c};\boldsymbol{\rho },\boldsymbol{0%
})|0,0\rangle $ is governed by the leading terms in the limit $\rho
\rightarrow 0$. Using the facts%
\begin{equation}
\delta \left( \boldsymbol{\rho }-\boldsymbol{\rho }^{\prime }\right) =\int d%
\boldsymbol{k}\frac{e^{i\boldsymbol{k}\cdot \left( \boldsymbol{\rho }-%
\boldsymbol{\rho }^{\prime }\right) }}{\left( 2\pi \right) ^{2}}\left( \sum_{%
\boldsymbol{\alpha }}|\boldsymbol{\alpha }\left( \boldsymbol{q},\boldsymbol{k%
}\right) \rangle \langle \boldsymbol{\alpha }\left( \boldsymbol{q},%
\boldsymbol{k}\right) |\right)
\end{equation}%
and%
\begin{equation}
H_{0}^{\left( \mathrm{2D}\right) }\left( e^{i\boldsymbol{k}\cdot \boldsymbol{%
\rho }}|\boldsymbol{\alpha }\left( \boldsymbol{q},\boldsymbol{k}\right)
\rangle \right) =\varepsilon _{c}\left( e^{i\boldsymbol{k}\cdot \boldsymbol{%
\rho }}|\boldsymbol{\alpha }\left( \boldsymbol{q},\boldsymbol{k}\right)
\rangle \right) ,
\end{equation}%
it is easy to show that
\begin{eqnarray}
&&\langle 0,0|g(\varepsilon _{c};\boldsymbol{\rho },\boldsymbol{0}%
)|0,0\rangle   \notag  \label{ggga} \\
&=&\sum_{\boldsymbol{\alpha }^{\prime }}\int d\boldsymbol{k}^{\prime }\frac{%
e^{i\boldsymbol{k}^{\prime }\cdot \boldsymbol{\rho }}}{\left( 2\pi \right)
^{2}}\frac{\left\vert \langle 0,0|\boldsymbol{\alpha }^{\prime }\left(
\boldsymbol{q},\boldsymbol{k}^{\prime }\right) \rangle \right\vert ^{2}}{%
\varepsilon _{c}+i0^{+}-\varepsilon _{c^{\prime }}}  \notag \\
&=&\int d\boldsymbol{k}^{\prime }\frac{e^{i\boldsymbol{k}^{\prime }\cdot
\boldsymbol{\rho }}}{\left( 2\pi \right) ^{2}}\frac{1}{\varepsilon
_{c}+i0^{+}-|\boldsymbol{k}^{\prime }|^{2}}  \notag \\
&&+\sum_{\boldsymbol{\alpha }}\int d\boldsymbol{k}^{\prime }\frac{e^{i%
\boldsymbol{k}^{\prime }\cdot \boldsymbol{\rho }}}{\left( 2\pi \right) ^{2}}%
\left\vert \langle 0,0|\boldsymbol{\alpha }\left( \boldsymbol{q},\boldsymbol{%
k}^{\prime }\right) \rangle \right\vert ^{2}\times   \notag \\
&&\left( \frac{1}{\varepsilon _{c}+i0^{+}-\varepsilon _{c^{\prime }}}-\frac{1%
}{\varepsilon _{c}+i0^{+}-|\boldsymbol{k}^{\prime }|^{2}}\right) .  \notag \\
&&
\end{eqnarray}%
with $c^{\prime }=(\boldsymbol{\alpha }^{\prime };\boldsymbol{q};\boldsymbol{%
k}^{\prime }).$ Using the fact%
\begin{equation}
\int d\boldsymbol{k}^{\prime }\frac{e^{i\boldsymbol{k}^{\prime }\cdot
\boldsymbol{\rho }}}{\left( 2\pi \right) ^{2}}\frac{1}{\varepsilon
_{c}+i0^{+}-|\boldsymbol{k}^{\prime }|^{2}}=-\frac{K_{0}\left( -i\sqrt{%
\varepsilon _{c}}\rho \right) }{2\pi }
\end{equation}%
with $K_{0}$ the modified Bessel function, we get the result%
\begin{eqnarray}
&&\langle 0,0|g(\varepsilon _{c};\boldsymbol{\rho },\boldsymbol{0}%
)|0,0\rangle   \notag \\
&=&-\frac{1}{2\pi }K_{0}\left( -i\sqrt{\varepsilon _{c}}\rho \right)   \notag
\\
&&+\sum_{\boldsymbol{\alpha }}\int d\boldsymbol{k}^{\prime }\frac{e^{i%
\boldsymbol{k}^{\prime }\cdot \boldsymbol{\rho }}}{\left( 2\pi \right) ^{2}}%
\left\vert \langle 0,0|\boldsymbol{\alpha }\left( \boldsymbol{q},\boldsymbol{%
k}^{\prime }\right) \rangle \right\vert ^{2}\times   \notag \\
&&\left( \frac{1}{\varepsilon _{c}+i0^{+}-\varepsilon _{c^{\prime }}}-\frac{1%
}{\varepsilon _{c}+i0^{+}-|\boldsymbol{k}^{\prime }|^{2}}\right) .  \notag \\
&&
\end{eqnarray}%
In the small-distance region we have%
\begin{equation}
-\frac{1}{2\pi }K_{0}\left( -i\sqrt{\varepsilon _{c}}\rho \right) \approx
\frac{\ln \rho \!+\!C+\!\ln \left( -\frac{i\sqrt{\varepsilon _{c}}}{2}%
\right) }{2\pi }
\end{equation}%
and
\begin{eqnarray}
&&\sum_{\boldsymbol{\alpha }}\int d\boldsymbol{k}^{\prime }\frac{e^{i%
\boldsymbol{k}^{\prime }\cdot \boldsymbol{\rho }}}{\left( 2\pi \right) ^{2}}%
\left\vert \langle 0,0|\boldsymbol{\alpha }\left( \boldsymbol{q},\boldsymbol{%
k}^{\prime }\right) \rangle \right\vert ^{2}\times   \notag \\
&&\left( \frac{1}{\varepsilon _{c}+i0^{+}-\varepsilon _{c^{\prime }}}-\frac{1%
}{\varepsilon _{c}+i0^{+}-|\boldsymbol{k}^{\prime }|^{2}}\right)   \notag \\
&\approx &\lambda (\varepsilon _{c},\boldsymbol{q}).
\end{eqnarray}%
with $\lambda $-function defined in Eq.~(\ref{lam2d}). Then we have proved
Eq.~(\ref{gg}).

Eq.~(\ref{2}) can be proved with the similar approach. In the small-distance
region $r_{\ast }\lesssim r<<1/k$, the function $\langle 0,0|G(\varepsilon
_{c};\boldsymbol{\rho },\boldsymbol{0})|0,0\rangle $ is governed by the
leading terms in the limit $r\rightarrow 0$. Using the facts%
\begin{eqnarray}
\delta \left( \boldsymbol{r}-\boldsymbol{r}^{\prime }\right) &=&\int d%
\boldsymbol{k}\frac{e^{i\boldsymbol{k}\cdot \left( \boldsymbol{\rho }-%
\boldsymbol{\rho }^{\prime }\right) }}{\left( 2\pi \right) ^{2}}\left( \sum_{%
\boldsymbol{\alpha }}|\boldsymbol{\alpha }\left( \boldsymbol{q},\boldsymbol{k%
}\right) \rangle \langle \boldsymbol{\alpha }\left( \boldsymbol{q},%
\boldsymbol{k}\right) |\right)  \notag \\
&&\times \left( \sum_{n}\varphi _{n_{z}}\left( z\right) \varphi
_{n_{z}}\left( z^{\prime }\right) \right)
\end{eqnarray}%
and%
\begin{eqnarray}
&&\left( H_{0}^{\left( \mathrm{2D}\right) }+H_{z}\right) \varphi
_{n_{z}}\left( z\right) e^{i\boldsymbol{k}\cdot \boldsymbol{\rho }}|%
\boldsymbol{\alpha }\left( \boldsymbol{q},\boldsymbol{k}\right) \rangle
\notag \\
&=&\left( n\omega +\varepsilon _{c}\right) \varphi _{n_{z}}\left( z\right)
e^{i\boldsymbol{k}\cdot \boldsymbol{\rho }}|\boldsymbol{\alpha }\left(
\boldsymbol{q},\boldsymbol{k}\right) \rangle ,
\end{eqnarray}%
it is easy to prove that%
\begin{eqnarray}
&&\lim_{r\rightarrow 0}\langle 0,0|G(\varepsilon _{c};\boldsymbol{r},%
\boldsymbol{0})|0,0\rangle  \notag \\
&=&\lim_{r\rightarrow 0}\frac{1}{\varepsilon _{c}+i0^{+}-\left[
-\sum_{\alpha =x,y}\frac{\partial ^{2}}{\partial \alpha ^{2}}+H_{z}\right] }%
\delta \left( \boldsymbol{r}\right)  \notag \\
&&+\sum_{n_{z}=0}^{\infty }\!\left\vert \varphi _{n_{z}}\!\!\left( 0\right)
\right\vert ^{2}\!\!\lambda (\varepsilon _{c}-n_{z}\omega ;\boldsymbol{q})
\label{1b}
\end{eqnarray}%
As shown in Ref.~\cite{petrov01}, we have%
\begin{eqnarray}
&&\lim_{r\rightarrow 0}\frac{1}{\varepsilon _{c}+i0^{+}-\left[ -\sum_{\alpha
=x,y}\frac{\partial ^{2}}{\partial \alpha ^{2}}+H_{z}\right] }\delta \left(
\boldsymbol{r}\right)  \notag \\
&=&-\frac{1}{4\pi }\frac{1}{r}-\frac{w\left( \varepsilon _{c}/2\right) }{%
2\left( 2\pi \right) ^{3/2}l_{0}}  \label{1a}
\end{eqnarray}%
with the function $w\left( \eta \right) $ defined in Eq.~(\ref{w}).
Substituting Eqs~(\ref{1a}) into Eq. (\ref{1b}), we can get Eq.~(\ref{2}).

\section{Scattering amplitudes in 2D and quasi-2D geometries with SOC}

In this appendix we proof Eqs.~(\ref{t2}) and (\ref{feff}) for the 2D and
quasi-2D scattering amplitude. The 2D scattering amplitude $f^{\left(
\mathrm{2D}\right) }\left( c^{\prime }\leftarrow c\right) $ between the
incident state $|\psi _{c}^{(0)}\left( \boldsymbol{\rho }\right) \rangle $
and out-put state $|\psi _{c^{\prime }}^{(+)}\left( \boldsymbol{\rho }%
\right) \rangle $ is defined as%
\begin{equation}
f^{\left( \mathrm{2D}\right) }\left( c^{\prime }\leftarrow c\right) =-2\pi
^{2}\!\!\int \!\!d\boldsymbol{\rho }\langle \psi _{c^{\prime }}^{(0)}\left(
\boldsymbol{\rho }\right) |\ V_{\mathrm{2D}}\!\left( \rho \right) |\psi
_{c}^{(+)}\left( \boldsymbol{\rho }\right) \rangle ,  \label{ssb}
\end{equation}%
In the region $\rho ^{\prime }\lesssim \rho _{\ast }$ we have%
\begin{equation}
\langle \psi _{c^{\prime }}^{(0)}\left( \boldsymbol{\rho }\right) |\approx
\langle \psi _{c^{\prime }}^{(0)}\left( \boldsymbol{0}\right) |.  \label{ssa}
\end{equation}%
Substituting Eq.~(\ref{ssa}) into (\ref{ssb}), we immediately obtain the
result in Eq.~(\ref{t2}):
\begin{equation}
f^{\left( \mathrm{2D}\right) }\left( c^{\prime }\leftarrow c\right) =-2\pi
^{2}\!\langle \psi _{c^{\prime }}^{(0)}\left( \boldsymbol{0}\right)
|0,0\rangle A\left( c\right) .
\end{equation}

Eq. (\ref{feff}) for the quasi-2D scattering amplitude $f^{\left( \mathrm{Q2D%
}\right) }$ can be proved in the same approach. The definition of $f^{\left(
\mathrm{Q2D}\right) }$ is
\begin{equation}
f^{\left( \mathrm{2D}\right) }\left( c^{\prime }\leftarrow c\right)
=\!\!\!-2\pi ^{2}\!\!\int \!\!d\boldsymbol{r}\langle \Psi _{c^{\prime
}}^{(0)}\left( \boldsymbol{r}\right) |\ V_{\mathrm{3D}}\!\left( r\right)
|\Psi _{c}^{(+)}\left( \boldsymbol{r}\right) \rangle .
\end{equation}%
Using Eqs. (\ref{kaka}) and (\ref{isq2d}), we immediately get Eq. (\ref{feff}%
).

\section{2D scattering amplitude without SOC}

In this appendix we derive the 2D scattering amplitude of two
distinguishable atoms in systems without SOC \cite{petrov01}. In this case
we have the incident wave function%
\begin{equation}
\psi _{\boldsymbol{k}}^{\left( 0\right) }\left( \boldsymbol{\rho }\right) =%
\frac{1}{2\pi }e^{i\boldsymbol{k}\cdot \boldsymbol{\rho }}
\end{equation}%
with scattering energy $\varepsilon =k^{2}$. As shown in Appendix A and Ref.~%
\cite{petrov01}, the scattered wave function $\psi _{\boldsymbol{k}}^{\left(
+\right) }\left( \boldsymbol{\rho }\right) $ and the 2D scattering amplitude
$f^{\left( \mathrm{2D}\right) }\left( \varepsilon \right) $ are given by%
\begin{equation}
\psi _{\boldsymbol{k}}^{\left( +\right) }\left( \boldsymbol{\rho }\right)
=\psi _{\boldsymbol{k}}^{\left( 0\right) }\left( \boldsymbol{\rho }\right)
+Ag_{R}(\varepsilon ;\boldsymbol{\rho },\boldsymbol{0});\left( \rho \gtrsim
\rho _{\ast }\right)  \label{psi2d}
\end{equation}%
and%
\begin{equation}
f^{\left( \mathrm{2D}\right) }\left( \varepsilon \right) =-\pi A.
\end{equation}%
Here, we have
\begin{eqnarray}
g_{R}(\varepsilon ;\boldsymbol{\rho },\boldsymbol{\rho }^{\prime })
&=&_{\perp }\langle \boldsymbol{\rho }|\frac{1}{\varepsilon
+i0^{+}+\sum_{\alpha =x,y}\frac{\partial ^{2}}{\partial \alpha ^{2}}}\delta
\left( \boldsymbol{\rho }-\boldsymbol{\rho }^{\prime }\right)  \notag \\
&=&-\frac{1}{2\pi }K_{0}\left( -i\sqrt{\varepsilon }\rho \right) ,
\label{gr}
\end{eqnarray}%
where $K_{0}$ is the modified Bessel function of the second type. The
coefficient $A$ can be determined by considering that $\psi \left(
\boldsymbol{\rho }\right) \propto \ln \rho -\ln d$ in the region $\rho
_{\ast }\lesssim \rho \ll 1/\sqrt{\varepsilon }$, with $d$ the
characteristic length. Using this condition and the expression (\ref{gr}) of
$g_{R}(\varepsilon ;\boldsymbol{\rho },\boldsymbol{0})$, we get the
coefficient $A$ and then the 2D scattering amplitude $f_{0}^{\left( \mathrm{%
2D}\right) }$:%
\begin{equation}
f_{0}^{\left( \mathrm{2D}\right) }\left( \varepsilon \right) =\frac{\pi \psi
_{\boldsymbol{k}}^{\left( 0\right) }\left( \boldsymbol{0}\right) }{i\frac{1}{%
4}-\frac{1}{2\pi }\left( \ln d+C+\ln \frac{\sqrt{\varepsilon }}{2}\right) },
\label{f2dn}
\end{equation}%
where $C=0.5772...$ is the Euler gamma number.

The expression (\ref{f2dn}) clearly describes the character of the 2D
low-energy scattering. It shows that the low-energy scattering amplitude $%
f_{0}^{\left( \mathrm{2D}\right) }\left( \varepsilon \right) $ is totally
determined by the parameter $d$. In particular, $f_{0}^{\left( \mathrm{2D}%
\right) }\left( \varepsilon \right) $ logarithmically decays to zero~\cite%
{petrov01} in the limit $\varepsilon \rightarrow 0$, and achieves the
maximum when $\varepsilon =4\exp \left( -2C\right) /d^{2}.$

\section{Renormalization of the effective 2D interaction}

In this appendix we prove Eqs.~(\ref{re1}, \ref{re2}) and (\ref{re3}) for
the renormalization of the effective 2D interaction potentials $\hat{V}_{o}$
and $\hat{V}_{b}$. For the convenience of our calculation, in this appendix
we do not work in the $\boldsymbol{\rho }$-representation as before. We use
the Dirac vector $|\rangle _{\perp }$ to describe the quantum state of the
2D spatial relative motion of the two atoms, and $|\rangle _{T}$ for the
total quantum state for both the relative motion and spin of the two atoms.
Namely, the total quantum $|\psi \rangle _{T}$ and the spinor wave function $%
|\psi \left( \boldsymbol{\rho }\right) \rangle $ defined in Eq. (1) is
related as
\begin{equation}
|\psi \rangle _{T}=\int d\boldsymbol{\rho }|\psi \left( \boldsymbol{\rho }%
\right) \rangle |\boldsymbol{\rho }\rangle _{\perp },
\end{equation}%
where $|\boldsymbol{\rho }\rangle _{\perp }$ is the eigen-state of the
atomic relative coordinate.

We first consider the single-channel potential $\hat{V}_{o}$ defined in Eq.~(%
\ref{vo}). To obtain the correct expression for the parameter $g$, we should
calculate the 2D scattering amplitude given by $\hat{V}_{o}$. This
scattering amplitude can be obtained from the Lippmman-Schwinger equation
for the 2-body $T$-operator $\hat{T}_{o}$ with respect to $\hat{V}_{o}$:%
\begin{equation}
\hat{T}_{o}\left( E\right) =\hat{V}_{o}+\hat{V}_{o}\hat{g}\left( E\right)
\hat{V}_{o}\left( E\right) ,  \label{lsea}
\end{equation}%
where the Green's operator $\hat{g}\left( \eta \right) $ is defined as%
\begin{equation}
\hat{g}\left( \eta \right) =\frac{1}{\eta +i0^{+}\!-\!H_{0}^{\left( \mathrm{%
2D}\right) }\!}.
\end{equation}%
and satisfies $g\left( \eta ,\boldsymbol{\rho },\boldsymbol{\rho }^{\prime
}\right) =_{\perp }\langle \boldsymbol{\rho }|\hat{g}\left( \eta \right) |%
\boldsymbol{\rho }^{\prime }\rangle _{\perp }$ with $g\left( \eta ,%
\boldsymbol{\rho },\boldsymbol{\rho }^{\prime }\right) $ defined in Eq.~(\ref%
{g2d}). Eq. (\ref{lsea}) can be solved directly with the first-quantization
form of $\hat{V}_{o}$. We find that the on-shell element of $\hat{T}%
_{o}\left( E\right) $ has a separable expression%
\begin{equation}
_{T}\langle \psi _{c^{\prime }}^{\left( 0\right) }|\hat{T}_{o}\left(
\varepsilon _{c}\right) |\psi _{c}^{\left( 0\right) }\rangle _{T}=\mathcal{V}%
\left( c^{\prime }\right) \mathcal{U}\left( c\right) ,
\end{equation}%
where $|\psi _{c}^{\left( 0\right) }\rangle _{T}$ is defined in Eq.~(\ref%
{psi2db}) and we have $\varepsilon _{c}=\varepsilon _{c^{\prime }}$. The
functions $\mathcal{V}\left( c^{\prime }\right) $ and $\mathcal{U}\left(
c\right) $ can be obtained from Eq.~(\ref{lsea}). The 2D scattering
amplitude given by $V_{o}$ is
\begin{equation}
f_{o}\left( c^{\prime }\leftarrow c\right) =-2\pi^{2}\ _{T}\langle \psi
_{c^{\prime }}^{\left( 0\right) }|\hat{T}_{o}\left( \varepsilon _{c}\right)
|\psi _{c}^{\left( 0\right) }\rangle _{T}.
\end{equation}%
As shown in Sec. IV, the 2D scattering amplitude given by $V_{o}$ should be
the same as the quasi-2D scattering amplitude, and then we should have%
\begin{equation}
f_{o}\left( c^{\prime }\leftarrow c\right) =f^{\left( \mathrm{Q2D}\right)
}\left( c^{\prime }\leftarrow c\right)
\end{equation}%
with $f^{\left( \mathrm{Q2D}\right) }$ defined in Eq.~(\ref{feff}). This
requirement directly leads to Eq.~(\ref{re1}).

Now we consider the two-channel model $\hat{V}_{b}$ and prove the
renormalization relations~(\ref{re2}) and~(\ref{re3}) of $\hat{V}%
_{b}$. To this end, we should calculate the bound state $|\Psi _{b}^{\left(
\mathrm{eff}\right) }\rangle _{T}$, which satisfies%
\begin{equation}
|\Psi _{b}^{\left( \mathrm{eff}\right) }\rangle _{T}=\hat{g}\left(
E_{b}\right) \hat{V}_{b}|\Psi _{b}^{\left( \mathrm{eff}\right) }\rangle _{T}.
\label{eb}
\end{equation}%
Equation~(\ref{eb}) can also be solved directly via the first-quantization
form of $\hat{V}_{b}$. We can derive the energy $E_{b}$ as well as the
components of $|\Psi _{b}^{\left( \mathrm{eff}\right) }\rangle _{T}$ in both
the open and the close channels. As shown in Sec. IV, the energy $E_{b}$ of $%
|\Psi _{b}^{\left( \mathrm{eff}\right) }\rangle _{T}$ should be the same as
that of the quasi-2D bound state $|\hat{\Psi}_{b}\rangle _{T}$, and then
satisfies Eq.~(\ref{ebq2d}). The open-channel probability of $|\Psi
_{b}^{\left( \mathrm{eff}\right) }\rangle _{T}$ is the same as the
transverse-ground-state probability of $|\Psi _{b}\rangle _{T}$, which is
given in Eq.~(\ref{tab}). These requirements directly lead to Eqs.~(\ref{re2}%
) and (\ref{re3}).

\end{subappendices}

\end{document}